\documentclass[a4paper,fleqn,usenatbib]{mnras}
\usepackage{amsmath}
\usepackage{txfonts}
\usepackage{multicol}

\usepackage[T1]{fontenc}
\usepackage{ae,aecompl}

\usepackage{graphicx}	% Including figure files
\usepackage{grffile}
\usepackage{epsf}

\AtBeginDocument{}%
\begin{document}
%units:
	\newcommand{\msun}{M_{\odot}}
	\newcommand{\rsun}{M_{\odot}}
	\newcommand{\kms}{\, {\rm km\, s}^{-1}}
	\newcommand{\cm}{\, {\rm cm}}
	\newcommand{\gm}{\, {\rm g}}
	\newcommand{\erg}{\, {\rm erg}}
	\newcommand{\kel}{\, {\rm K}}
	\newcommand{\kpc}{\, {\rm kpc}}
	\newcommand{\mpc}{\, {\rm Mpc}}
	\newcommand{\seg}{\, {\rm s}}
	\newcommand{\kev}{\, {\rm keV}}
	\newcommand{\hz}{\, {\rm Hz}}
	\newcommand{\etal}{et al.\ }
	\newcommand{\yr}{\, {\rm yr}}
	\newcommand{\gyr}{\, {\rm Gyr}}
	\newcommand{\eq}{eq.\ }
	\newcommand{\amunit}{\msun {\rm AU^2/yr}}
	\def\arcsec{''\hskip-3pt .}
	
	% symbols used
	\def\gapprox{\;\rlap{\lower 3.0pt                       % approximately smaller
			\hbox{$\sim$}}\raise 2.5pt\hbox{$>$}\;}
	\def\lapprox{\;\rlap{\lower 3.1pt                       % approximately smaller
			\hbox{$\sim$}}\raise 2.7pt\hbox{$<$}\;}
	
	% special case for 4 plots in panels across the page
	\newcommand{\figsizeFour}{9.0cm}
	
	% widths for figures for single and double columns
	\newcommand{\figwidthSingle}{14.0cm}
	\newcommand{\figwidthDouble}{7.50cm}
	
	% ALTER these if/when change from single to double columns
	\newcommand{\figbig}{\figwidthSingle}
	\newcommand{\figsmall}{\figwidthDouble}

	\newcommand{\figappend}{\figwidthSingle}
	
	% own commands for ref equations
	\newcommand{\reqOne}[1]{equation~(\ref{#1})}
	\newcommand{\reqTwo}[2]{equations~(\ref{#1}) and~(\ref{#2})}
	\newcommand{\reqNP}[1]{equation~\ref{#1}}
	\newcommand{\reqTwoNP}[2]{equations~\ref{#1} and~\ref{#2}}
	\newcommand{\reqTo}[2]{equation~(\ref{#1})-(\ref{#2})}
	% equation based commands
	\newcommand{\rn}[1]{(\ref{#1})}
	\newcommand{\ern}[1]{equation~(\ref{#1})}
	\newcommand{\be}{\begin{equation}}
	\newcommand{\ee}{\end{equation}}
	\newcommand{\ff}[2]{{\textstyle \frac{#1}{#2}}}
	\newcommand{\ben}{\begin{enumerate}}
		\newcommand{\een}{\end{enumerate}}
% text color comments:
	\newcommand{\tred[1]}{\textcolor{red}{#1}}
	\newcommand{\tgreen[1]}{\textcolor{green}{#1}}

\title[Direct $N$-body simulation of the Galactic centre]
{Direct $N$-body simulation of the Galactic centre}

\author[Panamarev et al]
  {Taras Panamarev$^{1,2,3}$\thanks{Corresponding author email: panamarevt@gmail.com}\thanks{Fellow of the International Max Planck Research School for Astronomy
  		and Cosmic Physics at the University of Heidelberg (IMPRS-HD)}, Andreas Just$^1$, Rainer Spurzem$^{3,4,1}$\thanks{Research Fellow at Frankfurt Institute for Advanced Studies (FIAS)},\\
  		\newauthor Peter Berczik$^{3,5,1}$, Long Wang$^{6,7}$ and Manuel Arca Sedda$^{1}$\\
   \\
      $^1$ Zentrum f\"ur Astronomie der Universit\"at Heidelberg, Astronomisches Rechen-Institut, M\"onchhofstr. 12-14, 69120 Heidelberg, Germany \\
      $^2$ Fesenkov Astrophysical Institute, Observatory 23, 050020 Almaty, Kazakhstan \\
      $^3$ National Astronomical Observatories and Key Laboratory of Computational Astrophysics, Chinese Academy of Sciences, 20A Datun Rd.,\\ Chaoyang District, 100012, Beijing, China\\
      $^4$ Kavli Institute for Astronomy and Astrophysics at Peking University, 5 Yiheyuan Rd., Haidian District, 100871, Beijing, China\\
      $^5$ Main Astronomical Observatory, National Academy of Sciences of Ukraine, 27 Akademika Zabolotnoho St., 03680, Kyiv, Ukraine\\
      $^6$ Argelander-Institut f\"ur Astronomie, Auf dem H\"ugel 71, 53121, Bonn, Germany\\
      $^7$ RIKEN Center for Computational Science, 7-1-26 Minatojima-minami-machi, Chuo-ku, Kobe, Hyogo 650-0047, Japan
      }

\maketitle

\begin{abstract}

We study the dynamics and evolution of the Milky Way nuclear star cluster performing a high resolution direct one-million-body simulation. Focusing on the interactions between such stellar system and the central supermassive black hole, we find that different stellar components adapt their overall distribution differently. After 5 Gyr, stellar mass black holes are characterized by a spatial distribution with power-slope $-1.75$, fully consistent with the prediction of Bahcall-Wolf pioneering work. Using the vast amount of data available, we infer the rate for tidal disruption events, being $4 \times 10^{-6}$ per yr, and estimate the number of objects that emit gravitational waves during the phases preceding the accretion onto the super-massive black hole, $\sim 270$ per Gyr. We show that some of these sources could form extreme mass-ratio inspirals. We follow the evolution of binary stars population, showing that the initial binary fraction of $5\%$ drops down to $2.5\%$ inside the inner parsec. Also, we explored the possible formation of binary systems containing a compact object, discussing the implications for millisecond pulsars formation and the development of Ia Supernovae.

\end{abstract}


%###########################################################################
\section{Introduction}\label{sec:INTRO}
%###########################################################################

Observations suggest that the Milky Way (MW) nucleus hosts the closest known supermassive black hole (SMBH; e.g. \citealt{Schoedel2002, Ghez2003a}). The proximity of the SMBH allows us to monitor individual stars in its immediate vicinity. Although our line-of-sight to the Galactic centre (GC) is obscured by dust clouds, the light absorption is sufficiently low in the near-infrared to give us a chance to follow the trajectories of individual stars from ground based telescopes using adaptive optics techniques. Such observations, that are being carried out for more than 25 years \citep{Ghez2005,Gillessen2009,Gillessen2017}, revealed the presence of a cluster of young (<100 Myr old) massive stars in the inner arcsecond (0.04 pc) of the MW, the so-called S-star cluster. Among them, the so-called S2 star has a period of only 16 years, thus it has already completed a full orbit since the beginning of its monitoring \citep{Schoedel2002, GRAVITY2018}. The features of stellar orbital motion in the S-star cluster represent the strongest evidence for the presence of a SMBH in
the Galaxy centre so far (see for instance \citealt{Ghez2000}; \citealt{Eckart2017} for a review). The S-star cluster is surrounded by a disc of even younger (<6 Myr old) and heavier stars \citep{LevinBeloborodov2003,PaumardEtAl2006, Yelda2014}. Such young populations are immersed in an extended and dense nuclear star cluster (NSC, see \citealt{Genzel2010} for a review) mostly comprised of old stars, with ages $\sim 10$ Gyr. Therefore, the Galactic Centre can be schematised as a multi-facet system, comprised of a central SMBH with mass $M_{\rm SMBH} = 4.3\times 10^6~{\rm M}_\odot$ and a young population of massive stars (the S-stars and the nuclear disc) surrounded by an old NSC with mass $M_\mathrm{NSC} \simeq 2.5 \times 10^7 \msun$ \citep{Schodel2014}. The inner part of the NSC features distinct dynamical components such as the S-star cluster and the disc of young massive stars.

%\citep[see][for a review]{Genzel2010}

The SMBH dominates stellar dynamics within a typical radius, called influence radius $r_{\rm inf}$, which encompasses the region where the SMBH potential equals the overall gravitational field of NSC stars (see reviews by \citealt{Alexander2005,Alexander2017}). As a result of orbital evolution, spatial distribution of stars within $r_{\rm inf}$ is expected to evolve toward a cusp distribution, being described by a power-law $\rho(r)\propto r^{-\gamma}$. In the case of a single mass component, \cite{Bahcall1976} showed that over the nucleus relaxation time, the $\gamma$ value approaches a limiting value $\gamma_{\rm BW} \simeq -1.75$. Early observations of the GC did not find good matching with the theory, as the spatial distribution of old stars seem to be flat, or even decreasing, in the inner 0.1 pc \citep[e.g.][]{Buchholz2009}. However, recent studies supported by both observations and numerical modelling alleviated the discrepancy \citep{Gallego-Cano2018,Schodel2018,Baumgardt2018}
  
If a star reaches a region where the SMBH tidal forces exceed its self gravity, it can undergo tidal disruption \citep{Hills1975,FR1976}. Such process, called tidal disruption event, can be observed via emission of the stellar debris, which gets heated while falling toward the event horizon. The classical solution of the mass fallback rate follows a power-law decay $\dot{M} \sim t^{-5/3}$ \citep{Rees1988,Phinney1989}. More than 20 tidal disruption events (TDEs) have been observed in other galaxies \citep{Komossa2015} implying a rate of $\sim 10^{-5} \mathrm{yr}^{-1} \mathrm{gal}^{-1}$ \citep{Stone2016}. The proximity of the TDEs to the event horizon of the SMBHs allows to test general relativity in the strong gravity regime.

In the case of compact stellar remnants, such as white dwarfs (WDs), neutron stars (NSs) and black holes (BHs), the accretion onto the SMBH will radiate the binding energy in form of low-frequency gravitational waves. As the compact stellar object approaches the last stable orbit, the emission of gravitational radiation becomes more efficient and it can be detected by space-borne interferometers like LISA \citep{Babak2017}. The inspiraling objects can make $\sim 10^3-10^5$ orbital revolutions before being swallowed by the SMBH. The analysis of such a signal will allow to obtain information on the space-time geometry and to measure the redshifted mass and spin of the SMBH with high accuracy \citep{Amaro-Seoane2007,Amaro-Seoane2015}.

The SMBH plays an important role also in shaping the evolution of binary stars affecting the mechanisms that regulate their formation and disruption. When a binary star approaches the SMBH it can be disrupted \citep{Hills1988}. A possible consequence of such interaction is that one component is captured by the SMBH and the second one is kicked out with a high velocity, up to several thousand $\kms$. Therefore, unveiling the origin of hypervelocity stars can provide useful information on the existence of the Galactic SMBH. We refer to the review provided recently by \cite{Brown2015} for further details. In general, binaries do not dominate the energy budget of the NSC because single stars bound to the SMBH can become very energetic \citep{Trenti2007}. The diverging velocity dispersion profile with decreasing radius from the SMBH implies that a \textit{hard} binary at outskirts of the NSC can become \textit{soft} near the centre and be disrupted by interactions with high-velocity single stars \citep{Hopman2009b}. Compact objects with a `normal' companion can form X-ray binaries. Recently, a growing number of observations revealed an overabundant presence of X-ray binaries at the GC \citep{Muno2005, Perez2015, Mori2015, Hailey2018, Zhu2018}, which might be connected with the GC formation history \citep{Arca-Sedda2017}. Binary dynamics can lead to the formation of millisecond pulsars, comprised of rapidly rotating pulsars spun up by its companion. These sources are expected to be the main reason responsible for the Gamma-ray excess observed at the GC, although other possibilities have been invoked \citep{Daylan2016, fermi17}, possibly related to the NSC formation history \citep{Brandt2015, Arca-Sedda2017, Fragione2017, Abbate2018}. Moreover, the possible detection of a pulsars population in the SMBH close vicinity can be crucial to probe general relativity in the strong field regime \citep{Psaltis2016}.

%As discussed above, the NSC represents a highly complex stellar system, thus a reliable modelling of such environment requires highly precise and detailed numerical simulations, which are on the other hand, extremely time-consuming. The fastest way of modelling star clusters with a central massive BH is to use Monte Carlo or Fokker-Planck approaches, but these methods are approximate \citep[e.g.][]{Spurzem1999}. In order to achieve high accuracy, direct $N$-body simulations are required. Pioneering work of modelling multi-mass stellar dynamics around a massive BH (intermediate mass BH in a globular cluster) via direct $N$-body simulations was done by \cite{BaumgardtEtAl2004b} and (for the case of galactic nuclei) by \citet{Freitag2006}. 
One of the latest direct $N$-body models of the GC was performed by \citet{Baumgardt2018}, hereafter BAS2018. The possibility to use more than 1 million bodies to model a galaxy centre becomes possible only in recent times \citep{Arca-Sedda2015, Arca-Sedda2017a, Arca-Sedda2017b}. However, most of the existing models in the literature did not include all the relevant features at the same time, like a sufficiently large number of bodies, stellar evolution or a proper treatment for close encounters. Recently, the first realistic star-by-star simulations were performed for globular clusters, the DRAGON simulations \citep{Wang2016}, where the authors were able to track the stellar evolution for single and binary stars. In this paper we apply a similar approach to the case of the MW NSC. 

The paper is organized as follows. In Section~\ref{sec:method} we describe the method and initial conditions. Section~\ref{sec:genev} focuses on the evolution of the stellar system in general. Interactions of the system with the SMBH as well as the TDEs are described in Section~\ref{sec:bh}. In Section~\ref{sec:bin} we describe the evolution of compact binaries, whereas Section~\ref{sec:CON} is devoted to discuss and summarize our main results.

%###########################################################################
\section{Method}
\label{sec:method}
%###########################################################################

We model the NSC of the MW galaxy using the direct $N$-body fully parallel code NBODY6++GPU  \citep{Wang2015}. The code is a multi-node massively parallel extension of NBODY6 \citep{Aarseth2003} and NBODY6GPU \citep{Nitadori2012} and also features accurate treatment of binary stars and close encounters using the algorithm developed by \cite{KS1965} and the chain regularization \citep{Mikkola1993}. We refer to \citet{Wang2015} for a detailed description of the numerical features and differences with NBODY6/NBODY6GPU.

We approximate the NSC with $N \simeq 10^6$ particles. Although this is the largest number of particles ever used in the direct $N$-body modelling of the GC so far, the real number of stars in the MW NSC is up to two orders higher. The simulation takes into account stellar evolution as well as the formation and evolution of binary stars. We start the simulation after gas removal and after the onset of virial equilibrium. Moreover, we include in our model a population of initial binaries, being $5\%$ of the total particles number. Our choice is compatible with observational evidences suggesting that globular clusters dense cores are expected to host a low fraction of binaries \citep{Bellazzini2002b}. As shown by \cite{Ivanova2005} via numerical models, even assuming a $100\%$ fraction of initial binaries, a typical globular cluster would retain only 5-10 per cent of them at present-day. In NSCs, this fraction can be even lower, due to the higher velocity dispersion that tend to enhance binary disruption via close encounters \citep[e.g.][]{Hopman2009b}. Many of the initial binaries are wide and destroyed in the first few dynamical times by few-body interactions. From Sec. \ref{sec:bin} and Fig. \ref{fig:nbin} we conclude that the binary fraction of 5\% on average is quite stable with time; therefore we think that choosing initial binary fractions larger or even much larger than 5\% will not significantly affect the results for the long term evolution.

The SMBH is included as an external point-mass potential with initial mass of 10\% of the total stellar mass of the system (given the initial NSC mass of $4.0\times 10^7 M_\odot$). The SMBH can grow via stars accretion if a star's orbit intersect the region encompassed by the accretion radius, $r_{\rm acc}$. In our model, we assume $r_{\rm acc} = 4.2\times 10^{-4}{\rm ~pc} = 10^{3}R_S$, being $R_S$ the SMBH's Schwarzschild radius. In the case of an SMBH with mass $M_{\rm SMBH} = 4.3\times 10^6{\rm ~M}_\odot$, a Sun-like star undergoes disruption at a distance larger than the Schwarzschild radius, and it can appear observationally as a TDE. Compact remnants (WDs, BHs and NSs) disruption radius falls inside $R_S$ and their accretion onto the SMBH is likely not associated to any electromagnetic counterpart. On the other hand, compact remnants can tightly bind to the SMBH and undergo slow inspiral through low-frequency GWs emission. These so-called EMRIs represent a class of promising sources to be detected with space-borne detectors like LISA \citep[e.g.][]{Amaro-Seoane2007}, or TianQin \citep{TianQ2016}. We note that due to the limit of resolution in the number of particles we increase $r_t$ for all stars so that it equals to $r_\mathrm{acc}$\footnote{Note that the scheme we adopt for modelling stellar accretion leads the SMBH to consume more stars than in reality.}. In this simulation we analyse number counts for extreme mass ratio inspirals (accretion of compact objects onto the SMBH) as well as the TDEs by scaling $r_t$ and the number of events to real values as shown in Sec.~\ref{sec:tde}.

\subsection{Initial conditions}

To initialise our model, we construct a \cite{Plummer1911} equilibrium model immersed in a point-mass external potential (see \citealt{McMillan2007}), assuming $N=950k$ single stars and $N_b = 50k$ binary stars. The point-mass potential represents the SMBH which is fixed at the origin. We let the Plummer model adjust to the presence of the central potential and we start the modelling after this adjustment. Since this paper focuses on the inner part of the NSC, the effects from bulge, Galactic disc and dark matter halo are ignored.

We assumed a \cite{Kroupa2001} initial mass function, selecting masses in the range $0.08 - 100$ $\msun$. The initial binaries are paired with mass ratios $f(q) \propto q^{-0.4}$ motivated by observed values of the Scorpios OB2 association \citep{Kouwenhoven2007}, log-uniform distribution in semi-major axis with minimum and maximum values of 0.005  and 50 astronomical units\footnote{Since we want to cover the full possible range of binary parameters, some of them are overlapping at the initial moment. But they are merged at the next time-step, their number is very small and they do not affect the results discussed in this paper.} (AU) and thermal eccentricity distribution: $f(e) = 2e$.  

Single and binary stars are evolved using the stellar evolution packages SSE \citep{Hurley2000} and BSE \citep{Hurley2002a}. We assume that NSs at formation are subjected to a natal kick, whose amplitude is drawn according to a Maxwellian distribution with 1D velocity dispersion of $\sigma = 265 \kms$ \citep{Hobbs2005}. For BHs, the kick is calculated following the {\it fallback prescription} (see \citealt{Belczynski2002} for further details). The population of WDs, instead, is assumed to receive no kick at formation. The initial parameters are chosen to be as close as possible to those of the DRAGON simulations \citep{Wang2016}.

The star formation in the MW NSC is still ongoing and has complex history (see e.g. \citealt{Mapelli2016} for a recent review), but for simplicity 
we represent the NSC by a single stellar population of solar metallicity stars.

\subsection{Scaling}

In order to convert the original scale-free simulation of $10^6$ particles in $N$-body units to the real system in physical units, we assume the MW NSC mass to be $M_\mathrm{NSC} = 4.0\times 10^7 M_\odot = 10 M_\mathrm{SMBH}$. The \cite{Kroupa2001} initial mass function gives the simulated mass of $M_\mathrm{tot} = 6.18\times 10^5\msun$, thus one particle in the simulation represents a group of 65 stars. Therefore, stellar number counts are multiplied by 65 to be converted into real values. 

 For the radial scaling, we measure the influence radius of the SMBH at $t=0$ to be 0.66 $N$-body units and equate it to the value of the influence radius for the MW which is calculated using the central velocity dispersion taken from \cite{Gultekin2009}, $r_\mathrm{inf} = GM_\mathrm{SMBH}/\sigma^2 = 1.4$ $\mathrm{pc}$. Assuming that half-mass radius $r_\mathrm{hm} = 3r_\mathrm{inf}$, we can calculate the half-mass relaxation time \citep{Spitzer1987} 
\begin{equation}
t_\mathrm{rel} = \frac{0.14N}{\ln(0.4N)}\left(\frac{r_\mathrm{hm}^3}{GM_\mathrm{tot}}\right)^{1/2}
\end{equation}
for the MW NSC in physical units ($\simeq 11$ Gyr) and for the model in $N$-body units and scale the time accordingly. 
By equating the relaxation time of the model with the relaxation time of the real system we can set the stellar evolution time in correspondence with the dynamical time of the system by 
\begin{equation}
\frac{t_\mathrm{rel}'}{t_\mathrm{stev}'} = \frac{t_\mathrm{rel}}{t_\mathrm{stev}},
\end{equation}
where the prime denotes the modelled system and $t_\mathrm{stev}$ can be any stellar evolution time-scale.

The simulation was evolved up to 5.5 Gyr, which corresponds to a half of the initial half-mass relaxation time, but covers a few relaxation times inside the influence radius of the SMBH.

All values discussed below in the paper (densities, number counts, etc.) are given in physical units (except stated otherwise) for the \textit{realistic} MW NSC.

%###########################################################################
\section{General evolution of the system}
\label{sec:genev}
%###########################################################################

Throughout the simulation, the NSC lost roughly half of its initial mass mostly owing to stellar evolution with small contribution from the accretion of stars onto the SMBH (see Sec.\ref{sec:bh}). The final mass of the NSC is consistent with its present-day mass inferred from observations \citep{Schodel2014}.

\begin{figure}
\begin{centering}
\includegraphics[width=\columnwidth]{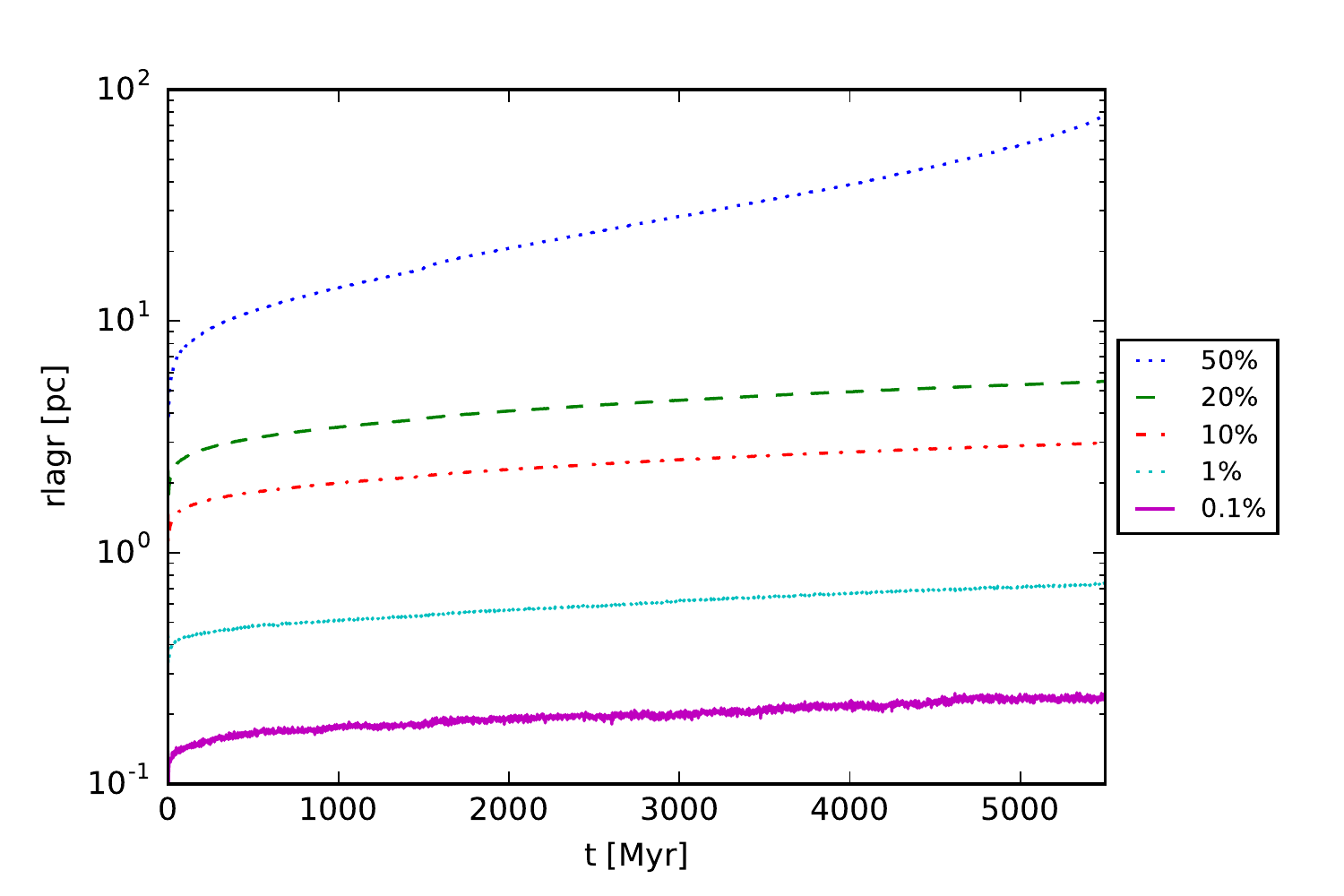} \\ 
\par\end{centering}
\caption{Evolution of the Lagrangian radii of the NSC. Blue dotted line shows the time evolution of 50\% Lagrange radius, while dashed green, dash-dotted red, dotted cyan and solid magenta lines correspond to 20, 10, 1 and 0.1\% respectively. }
\label{fig:rlagr}

\end{figure}

The NSC overall evolution can be monitored through the time evolution of the Lagrangian radii, which are the radii containing a certain fraction of the total stellar mass. As seen in Fig.~\ref{fig:rlagr}, the stellar system experiences an initial adjustment  to the SMBH potential, explained by the fact that the binaries are not taken into account for the generation of initial conditions. Also, there is a strong mass-loss rate during the first tens of Myr. Overall, the expansion of the NSC is driven by the stellar evolution mass-loss (the magenta line in Fig.~\ref{fig:rlagr} clearly shows the expansion), but the Galactic bulge would keep the outer Lagrange radii at roughly a constant value. The small expansion of the inner Lagrange radius (0.1\%) is driven by the accretion of stars onto the SMBH.	

\begin{figure}
\begin{centering}
\includegraphics[width=\columnwidth]{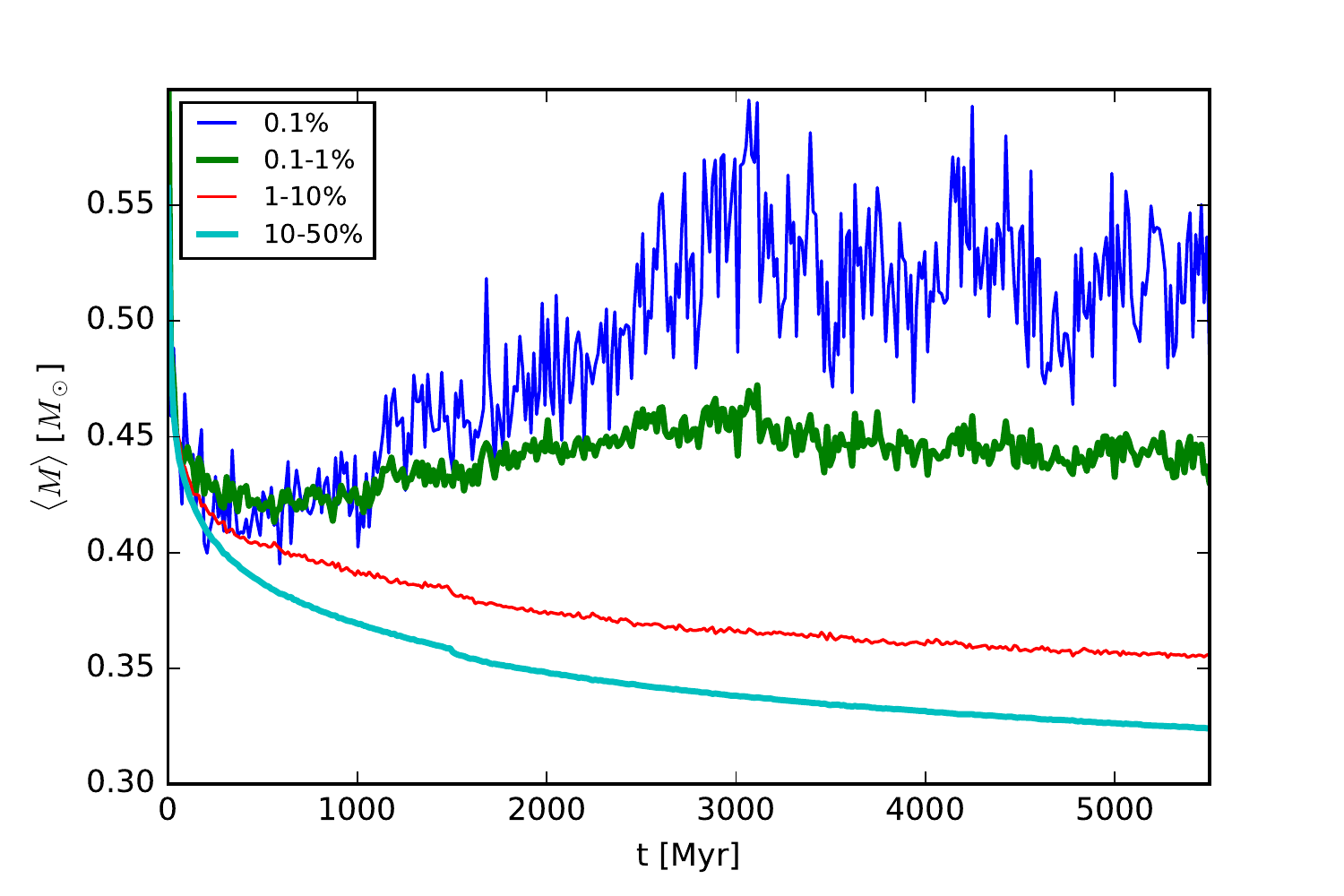} \\ 
\par\end{centering}
\caption{Time-evolution of the average stellar mass between shells of Lagrangian radii. Solid blue line corresponds to the evolution of average stellar mass in the region where Lagrangian radius is less than 0.1\%, thick green line shows the average mass between 0.1 and 1\%, red and cyan lines represent masses between 1 - 10\% and 10 - 50\% respectively. Two upper curves indicate mass segregation.}
\label{fig:mavg}
\end{figure}

\begin{figure*}
%\begin{centering}
	\includegraphics[width=0.8\linewidth]{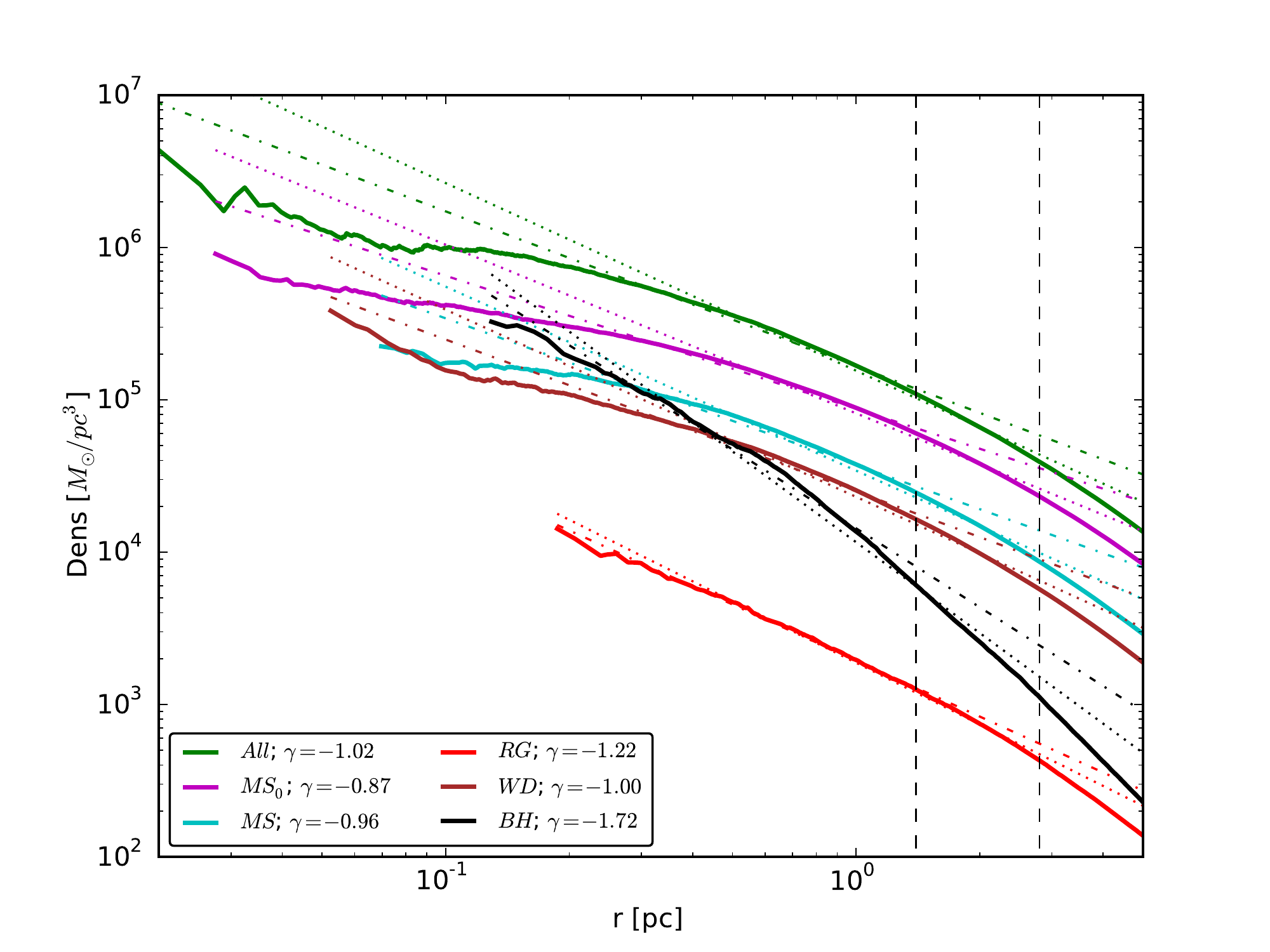}\par  
%	\par\end{centering}
\caption{Stellar density profiles at $t = 5$ Gyr for different stellar types. Thick solid lines correspond to: All - all stars, MS$_\mathrm{low}$ - low mass main sequence stars, MS - main sequence, RG - red giants, WD - white dwarfs, BH - black holes. Corresponding power-law slopes fitted inside the initial and final influence radii of the SMBH are shown as dash-dotted and dotted lines of the same colour. The dashed vertical lines denote the initial influence radius ($r=1.4$ $\mathrm{pc}$) and the influence radius at $t=5$ $\mathrm{Gyr}$ ($r\sim 2.8$ $\mathrm{pc}$) of the SMBH. The power-law indices fitted inside $r=1.4$ $\mathrm{pc}$ are shown in the legend.}
\label{fig:rho}
\end{figure*}

The time evolution of the average stellar mass, in Lagrangian shells, reveals mass segregation, as shown in Fig.~\ref{fig:mavg}. After all the heavy stars lost most of their mass ($\sim 300$ Myr), the mass segregation overtakes the time evolution of the average masses in Lagrangian shells. After $\sim 3$ Gyr of evolution, a quasi-steady state is established for the innermost regions. The total number of stars in terms of different stellar evolution components\footnote{The stellar types are defined as in \cite{Hurley2000}, but for simplicity we combine different types of RGs in one stellar type.} and their properties are described in subsequent sections and summarized in Table~\ref{StProp}.

\subsection{Density profiles}

Typically the 3D stellar density (as well as the surface density) is described by a power law of the form $\rho(r) \propto r^\gamma$, where $r$ is the distance from the SMBH. For the case of equal mass solar type stars the slope becomes $\gamma=-1.75$ inside the influence radius of the SMBH \citep{Bahcall1976}. For the case of a mass spectrum the dominant component obtains the -1.75 slope \citep{Bahcall1977}. 

In Fig.~\ref{fig:rho} we present 3D stellar density profiles for various stellar types. In order to get a better accuracy, we measured the density profile power-law slopes for 10 snapshots around $t = 5$ Gyr and averaged the results. Due to the low particle number in the inner part, we required at least 3 particles for the calculation of the density. Stellar mass BHs have the steepest slope of $\gamma=-1.72\pm0.04$ while the low mass and high mass main sequence (MS) stars are characterized by a shallower slope, being $\gamma=-0.87\pm0.01$ and $\gamma=-0.96\pm0.02$, respectively, calculated at 5 Gyr. The slopes are measured inside the SMBH influence radius. White dwarfs (WDs) have a similar slope ($\gamma=-1.00\pm0.02$), but red giants (RGs) are slightly steeper with $\gamma=-1.22\pm0.12$. Comparison with BAS2018 (see their Fig. 2) yields very similar slopes for the giants, although, for the upper and lower MS stars their simulations show steeper slopes. In principle, the results are consistent with each other since BAS2018 use slightly different definitions for lower and upper MS stars, and they show the results at $t = 13$ $\mathrm{Gyr}$. Another point is that BAS2018 have exponentially declining star formation rate, they implement it by adding new stars every\gyr. The power law slope for the BHs is remarkably consistent with the analytical prediction of \citet{Bahcall1977} and with a recent study based on Fokker-Planck approach \citep{Vasiliev2017}. The cusp is already formed at $t < 2$ $\mathrm{Gyr}$, less than one NSC half-mass relaxation time. However, as shown by \citet{AmaroPreto2011}, the cusp regrowth time is 1/4 of the relaxation time, thus our results are consistent with the assumed initial half-mass relaxation time. 

Due to stellar mass-loss the influence radius of the SMBH expanded from 1.4 to 2.8 pc and we present the linear fitting for the density slopes for the region $r$ < 2.8 pc as well (see columns 4-5 of Table~\ref{StProp}). The power-law indices for the influence radius at $5 \gyr$ are more consistent with the strong mass segregation solution, but are still shallower than the values proposed by \cite{Alexander&Hopman2009} and \cite{Preto&Seoane2010}. They claimed that in the case when the number of lower mass objects (stars with masses up to $1 M_\odot$) is much higher than that of heavy objects (stellar BHs with masses of $10 M_\odot $) the heavy objects obtain a power-law density slope $\gamma$ of -11/4 while the light ones have $\gamma = -3/2$. They parametrized the solution by $\Delta = N_hM_h^2/(N_lM_l^2) 4/(3+M_h/M_l)$, where $N$ and $M$ denote the numbers and masses of light and heavy objects. In our simulation the value of $\Delta$ approaches zero ($\Delta \sim 5\times 10^{-8}$) meaning that we are in the strong mass segregation regime, although the density slopes in our simulation are shallower than predictions.

As seen in Fig.~\ref{fig:rho}, low-mass MS stars dominate at $r > 0.1$ pc and low-number statistics at smaller radii does not allow us to study the details of stellar density distribution there. We leave this analysis for future work.    

\subsection{Stellar mass black holes and other compact objects}
\begin{figure}

    \includegraphics[width=\columnwidth]{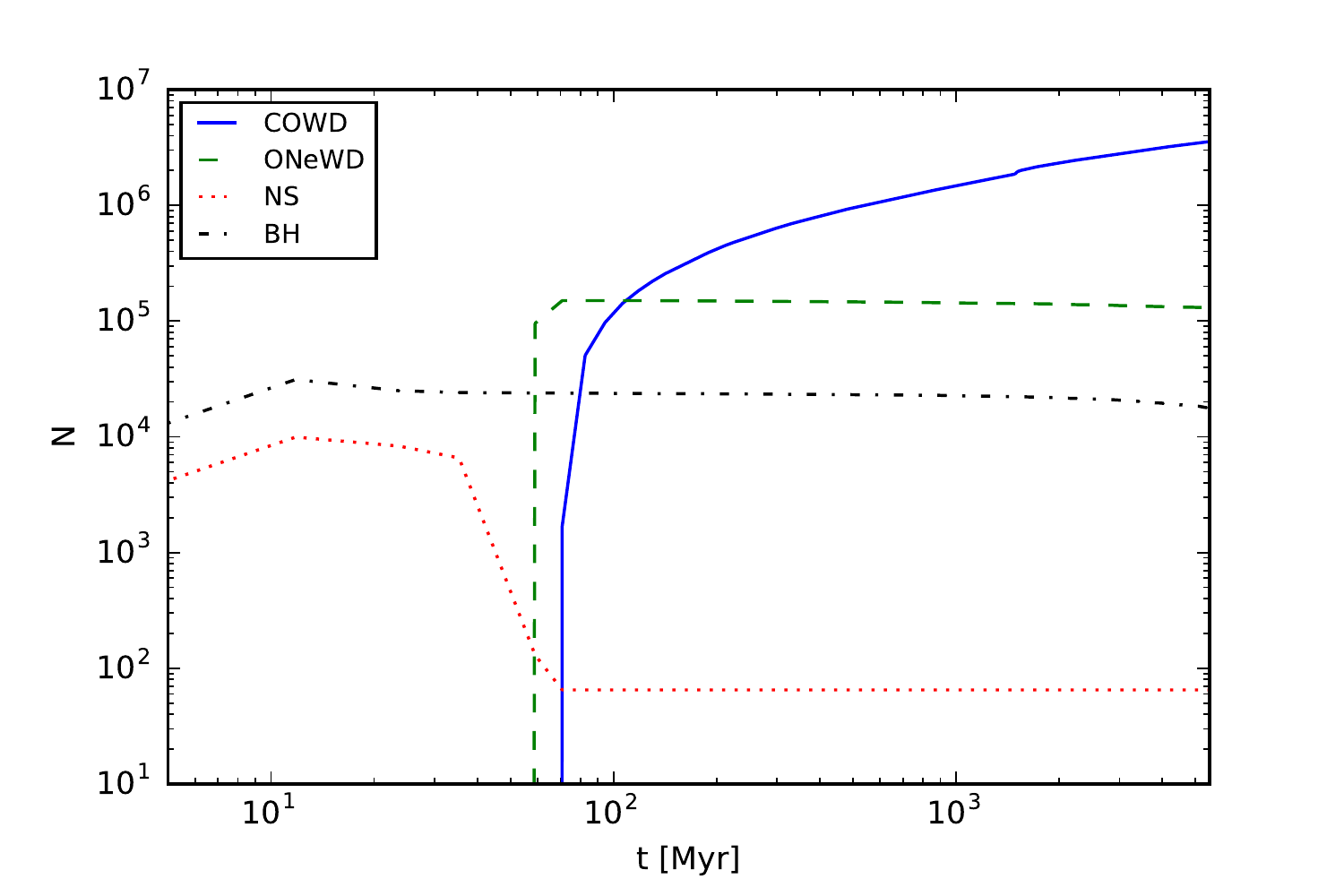}\par

\caption{The number of compact objects as function of time. All lines show only the single stars. Solid blue and dashed green lines show WDs, dash-dotted black and dotted red lines show BHs and NSs respectively. High natal kicks remove most of the NSs from the system, while there are still more than $10^4$ stellar BHs.}
\label{fig:SCDens}
\end{figure}

\begin{figure}
\begin{centering}
\includegraphics[width=\columnwidth]{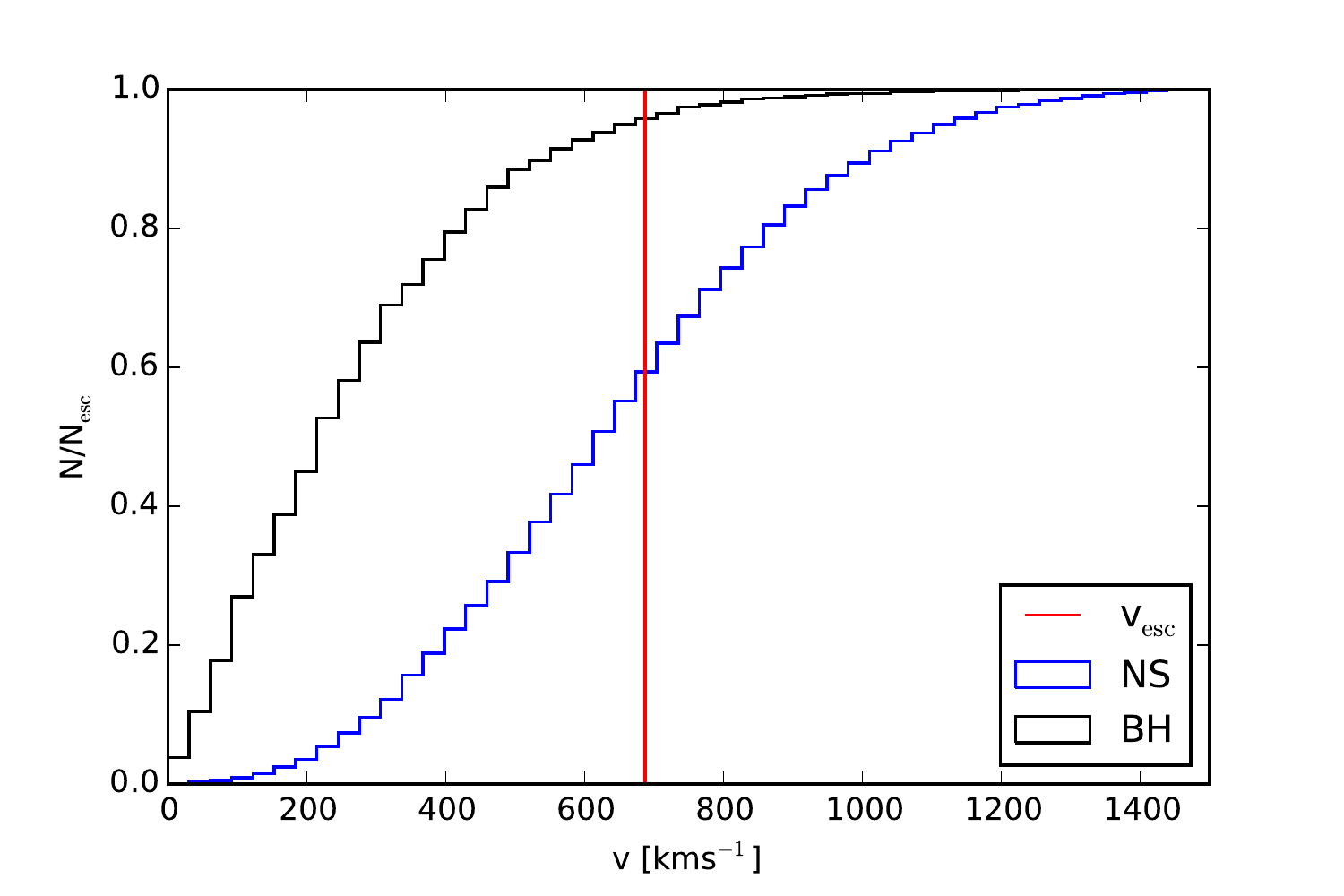} \\ 
\par\end{centering}
\caption{Cumulative normalized histogram showing velocities of the escaped NSs (blue) and BHs (black). The red vertical line represents the escape velocity for the MW bulge at 100 pc. } 
\label{fig:esc_nsbh}
\end{figure}

Compact objects may play an important role in the evolution of the NSC. Fig.~\ref{fig:SCDens} shows the time evolution of compact objects divided by type: carbon-oxygen white dwarfs (COWD), oxygen-neon white dwarfs (ONeWD), NSs and BHs. After 5 Gyr, the population of COWDs is still growing due to stellar evolution, while the formation of ONeWDs already ceased after 100 Myr, although
$\sim 1.4\times 10^5$ ($\sim 1.3\times 10^5$) of them are still retained at 2 (5) Gyr. While WDs represent still a noticeable population after 5 Gyr, almost all the NSs are ejected, due to the high natal kick received consequently to supernova explosions. We have to note that in a real galactic nucleus the NSs may be still bound to the system under the influence of the potential from the galactic bulge and dark matter halo, that becomes more important at the outer boundaries of the NSC. Fig.~\ref{fig:esc_nsbh} shows the normalized distribution of velocities for the escaped NSs and BHs calculated at 100 pc. Assuming that the MW bulge potential is reasonably represented by a standard Plummer sphere $\Phi = - GM_\mathrm{tot}/\sqrt{r^2+b^2}$ with total mass $M_\mathrm{tot} = 2.0\times10^{10} M_\odot $ \citep{Valenti2016} and the scale length $b = 350$ pc \citep{Dauphole1995}, we found that $\sim 60$\% of escaped NSs have velocities lower than the bulge escape velocity calculated at 100 pc. This suggests that as many as $8\times10^5$ NSs might be still wandering in the galactic bulge, and possibly can come back to the NSC. For the stellar-mass BH population the situation is different in a way that their kick velocity depends on the fallback factor \citep{Belczynski2002}. This explains the initial peak in the number of stellar BHs (dash-dotted line of Fig.~\ref{fig:SCDens}): more than half of them escaped but after that the number of BHs declines slowly and we expect $\sim 2.2\times 10^4$ ($\sim 1.8\times 10^4$) stellar-mass BHs after 2 (5) Gyr. The black line in Fig.~\ref{fig:esc_nsbh} shows that $\simeq95$\% of all escaped BHs would be still bound to the system, increasing their total number.  

\begin{figure}
\begin{centering}
\includegraphics[width=\columnwidth]{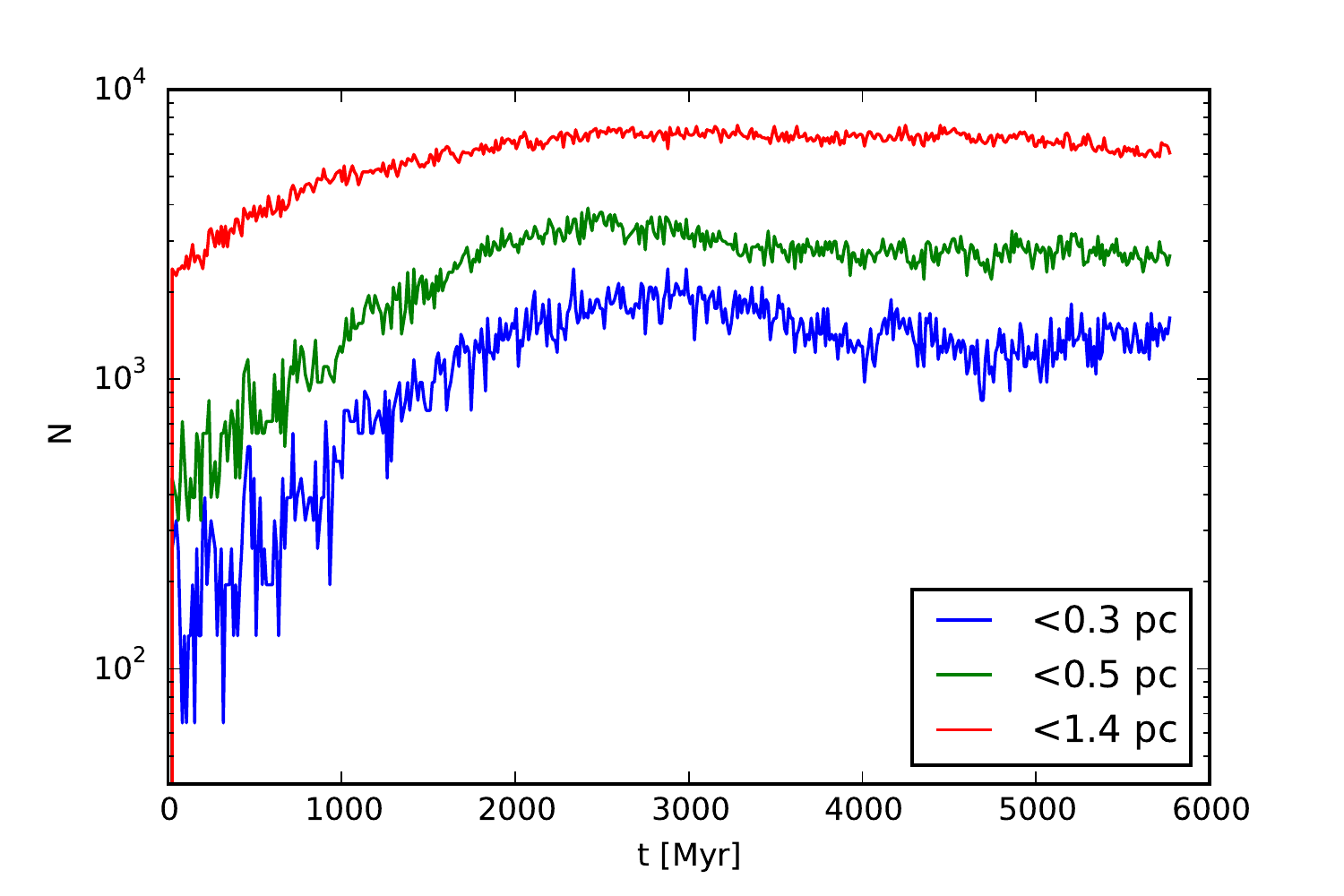} \\ 
\par\end{centering}
\caption{The number of BHs as function of time. The red, green and blue line show the number of stellar mass BHs inside 1.4, 0.5 and 0.3 pc respectively.}
\label{fig:nbh}
\end{figure}

Fig.~\ref{fig:nbh} shows the number of stellar-mass BHs in the inner part of the NSC as a function of time. Since the BHs are the heaviest objects in the NSC, they experience the strongest mass segregation. The number of BHs in the inner 0.5 and 0.3 pc increased significantly over time. We expect $\sim 2000$ and $\sim 1000$ BHs inside central 0.5 and 0.3 pc respectively and $\sim 6000$ inside the initial influence radius of the SMBH (1.4 pc) at $5$ Gyr. Having in mind Fig.  \ref{fig:esc_nsbh}, we note that the numbers of BHs above have to be treated as lower limits.

%###########################################################################
\section{The supermassive black hole}
\label{sec:bh}
%###########################################################################

Close stellar passages around the SMBH may result either in the star disruption, a phenomenon called TDE, or in its gravitational wave (GW) induced inspiral/plunge. In the latter case, the tight SMBH-star binary evolve mostly through GW emission, possibly resulting in  a so-called extreme mass ratio inspiral (EMRI).

In this section, we try to quantify the amount of TDEs and EMRIs expected to form over the whole NSC lifetime.

\subsection{Tidal disruption events}\label{sec:tde}

\begin{table*}
\caption{Properties of different stellar types at $t=5\gyr$.}
\label{StProp}
\begin{tabular}{llllllll}
\hline
Stellar type & $N_\mathrm{tot}$ & <$m$>  & $\gamma (r < 1.4 \mathrm{pc})$ & $\gamma (r < 2.8 \mathrm{pc})$ & <$r_t$> & $\dot{N}_\mathrm{acc}$ & $\dot{N}_\mathrm{EMRI} $  \\
 &  & ($\msun$) &  &  & pc & (Gyr$^{-1}$) & (Gyr$^{-1}$)\\
\hline
Low-mass main sequence  & $5.0\times10^7$ &  0.25 & $-0.87\pm0.01$ & $-1.11\pm0.03$ & $1.5\times10^{-6}$ & 3278 & - \\
Main sequence   & $4.3\times10^6$ &   0.92 & $-0.96\pm0.02$ & $-1.21\pm0.05$ & $4.0\times10^{-6}$ & 658 & - \\
Red giant   & $1.5\times10^5$ &   1.24 & $-1.22\pm0.12$ & $-1.34\pm0.15$ & $4.9\times10^{-5}$ & 39 & - \\
White dwarf       & $1.6\times10^6$ &   0.71 & $-1.00\pm0.02$ & $-1.23\pm0.03$ & $1.6\times10^{-6}$ & 255 & 2\\
Black hole  & $1.8\times10^4$ &  10.05 & $-1.72\pm0.04$ & $-1.98\pm0.07$ & $1.6\times10^{-6}$ & 4 & 2\\
All stars  & $5.8\times10^7$ &  0.33 & $-1.02\pm0.02$ & $-1.23\pm0.03$ & - & 4120  & -\\
\hline
\end{tabular}
\par\medskip
\begin{flushleft}\textbf{Notes.} Column 1 is the name of the stellar type, columns 2 and 3 show the total number of stars at 5 Gyr and the average stellar mass in solar masses, columns 4 and 5 represent the 3D density power-law indices inside 1.4 and 2.8 pc respectively, column 6 shows the average tidal disruption radius that was used in Eq.~\ref{nscale} (for compact objects we used the value of $4R_\mathrm{S}$ as described in Sec.~\ref{sec:gw}), column 7 gives the number accretion rate per Gyr derived over the period of 5 Gyr and column 8 shows the EMRI rate per Gyr for BHs and WDs. Numbers in columns 7 and 8 are rescaled using Eq.~\ref{nscale}. \end{flushleft}
\end{table*}

A star with mass $m_*$ and radius $r_*$ can get tidally disrupted if the SMBH tidal forces overcome the star self-gravity. The resulting stellar debris distribute in a disc, feeding the SMBH while emitting X-ray radiation, giving rise to an observable phenomenon called TDE \citep{Hills1975, FR1976}.
Equating the gradients of these two competing forces allows us to calculate the tidal disruption radius, which is given by % is given by
\begin{equation}
\label{rtid}
r_\mathrm{t} \simeq r_*\left(\frac{M_\mathrm{SMBH}}{m_*}\right)^{1/3}.
\end{equation}
In our model, we assumed that a star passing sufficiently close to the SMBH is completely accreted, without any mass left. Since the number of particles used to model the NSC is 65 times smaller than in the real NSC, the number of possible TDEs is limited by low-resolution in the SMBH vicinity. To deal with this problem, we initially set a large tidal radius, $r_t = 4.2\times 10^{-4}$ pc (the same for all objects), scaling down \textit{a posteriori} to the actual $r_t$ (computed using stellar radius and mass at the moment of accretion) values. In particular, we scale the number of events using the relation obtained from loss-cone theory, according to which the number of stars accreted through tidal disruption 
depend on the stellar tidal radius and the total number of stars in the system,
\begin{equation}
N_\mathrm{acc} \propto r_t^{4/9}\times \left(N/\ln(0.4N)\right)^{4/9}
\end{equation}
\citep{BaumgardtEtAl2004a, KenEtAl2016}.
Therefore, the number of accreted stars in the real system can be estimated using the above scaling relation
\begin{equation}
\label{nscale}
N_\mathrm{acc}^\mathrm{real} = \left(\frac{r_t^\mathrm{real}}{r_t^\mathrm{sim}}\right)^{4/9} \times \left(\frac{N_\mathrm{real}}{N_\mathrm{sim}}\right)^{4/9} \left( \frac{1}{\ln(0.4N_\mathrm{real}/N_\mathrm{sim})} \right)^{4/9} \times N_\mathrm{acc}^\mathrm{sim}.
\end{equation}

For each of the 5 star groups summarized in Table~\ref{StProp}, we calculated the corresponding average tidal radius through Eq.~\ref{rtid} using the values for stellar mass and radius at the moment of accretion (see column 7 of Table~\ref{StProp}), and the number of stars passing closer than $r_t$ in our simulation, namely $N_{\rm acc}^{\rm sim}$. This quantity is then used in Eq.~\ref{nscale} to scale our results to the real NSC.

\begin{figure}
	\begin{centering}
		\includegraphics[width=\columnwidth]{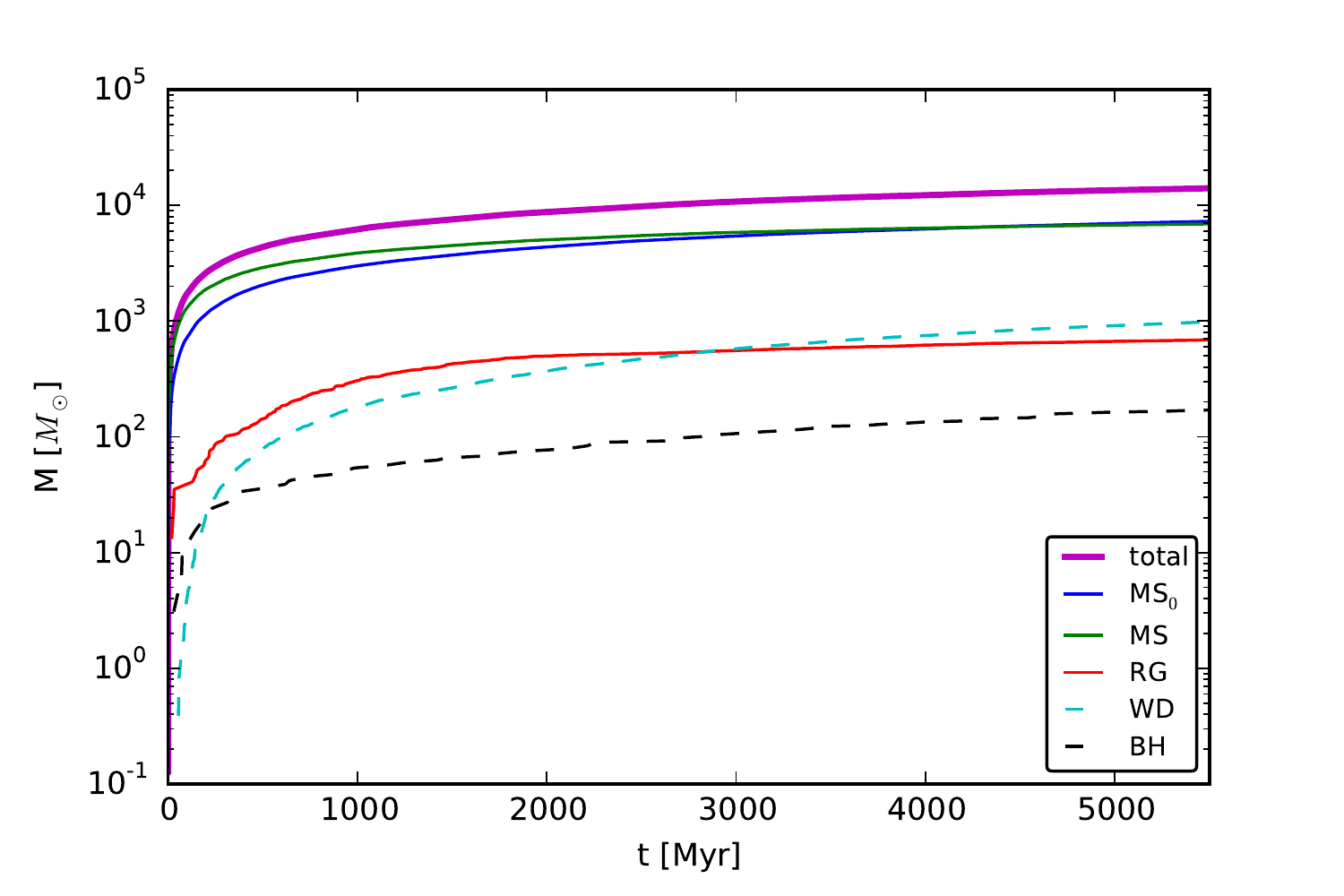} \\ 
		\par\end{centering}
	\caption{Growth of the SMBH. The thick magenta line represents the total accreted mass onto the SMBH, other lines the contributions from low mass MS stars (blue), MS stars (green), RGs (red), WDs (cyan) and BHs (black). The accreted mass is calculated using Eq. \ref{nscale}}.
	\label{fig:massbh}
\end{figure}

Table~\ref{StProp} (columns 7 and 8) lists the number of tidally disrupted (or accreted) stars per Gyr derived from the total number in 5 Gyr (scaled using Eq.\ref{nscale}). 
The majority of TDEs are due to low-mass MS stars, while the SMBH growth is mostly due to MS stars. As shown in Fig.~\ref{fig:massbh}, the time evolution of the accreted mass saturates to a nearly constant value in $2.8$ Gyr, allowing us to provide an upper limit to the SMBH accreted mass by 5 Gyr as $\Delta M_\mathrm{SMBH} \simeq 10^4 M_\odot$ or $\simeq 0.23\% $ of the initial SMBH mass. This implies a mass accretion rate $\dot{M} = 2.0\times 10^{-6} M_\odot \mathrm{yr^{-1}}$ and a TDE rate $\dot{N}_\mathrm{TDE} \simeq 4.1\times 10^{-6}$\yr$^{-1}$ which is consistent with the observed number of TDEs obtained per MW-like galaxy \citep{Stone2016}. We note that due to the initial loss-cone depletion the accretion rate is higher in the beginning and is smaller at later stages of evolution. 

\begin{figure}
	\begin{centering}
		\includegraphics[width=\columnwidth]{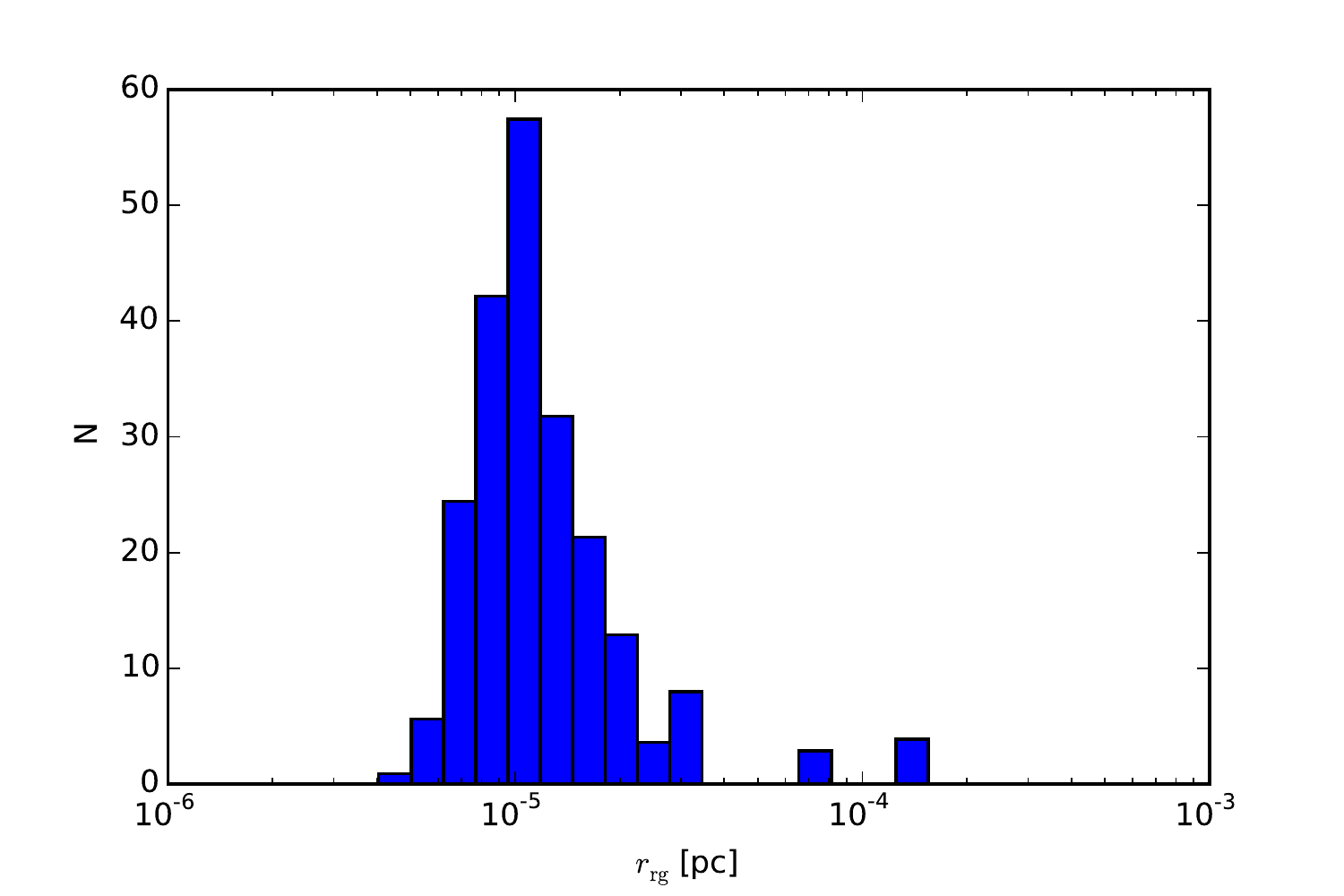} \\ 
		\par\end{centering}
	\caption{Histogram of the locations where the red giants were stripped by the SMBH.}
	\label{fig:rg_loc}
\end{figure}

As stated above, we assume that a star undergoing a TDE in our model is completely disrupted and 100\% of its mass is added to the SMBH. However, the tidal disruption radius for a RG is large while its typical density is generally low, thus a close encounter
with the SMBH may lead to an envelope stripping \citep{MacLeod2012, Bogdanovic2014} and the core will remain with the structure similar to a WD (e.g. \citealt{Althaus1997}). These WDs are very hot ($\sim 10^5 K$) and may be observable. Fig.~\ref{fig:rg_loc} shows possible locations of these objects, the radii at which the RGs were stripped (column 7 of Table~\ref{StProp} gives numbers of such events), but we do not follow the dynamics of the survived core after the disruption. The detection of the survived cores in the GC may give constrains on the number of giant disruptions. The remnant WDs may also increase the fraction of EMRIs (see next subsection), but these effects will be explored in future work.   

\subsection{Gravitational waves}\label{sec:gw}

In general, compact objects (WDs, NSs or BHs) can survive tidal disruption due to their compact sizes and can lead to the formation of an EMRI, a tight binary emitting GWs in the LISA \citep[e.g.][]{Amaro-Seoane2007} and TianQin \citep{TianQ2016} expected observational bands. In our models, we consider the accretion of a compact object if it gets inside the last parabolic stable orbit, $4 R_\mathrm{S}$ \citep{Amaro-Seoane2007, Amaro-Seoane2013}. Objects scattering directly into the $4 R_\mathrm{S}$ are direct plunges, they emit a burst of gravitational radiation, but are difficult to detect even for the GC since they do not spend much time in the LISA band. The number of direct plunges can be calculated using the loss-cone theory, applying the same procedure as in previous subsection. Column 7 of Table~\ref{StProp} lists the numbers for WDs and BHs which we classify as direct plunges. The subclass of these objects would 'plunge' into to the region between $1 R_\mathrm{S}$ and $4 R_\mathrm{S}$, we call them \textit{semi-plunges}. The semi-plunges may still make a couple of orbits (depending on the spin of the SMBH) and produce very different gravitational wave signal, but it is still difficult to detect them. We find the rate of semi-plunges to be 2 (117) for BHs (WDs) per Gyr. The electromagnetic counterpart will also be different for plunges and semi-plunges for the case of inspiralling WDs or NSs (see e.g. \citealt{Belczynski2018} where the authors study double NS mergers).

On another hand, EMRIs evolution is a process that the compact object undergoes a large number of close encounters with the SMBH \citep{Alexander&Hopman2003, Amaro-Seoane2007} that causes energy loss due to gravitational wave emission and, thus, may be detectable by LISA. To verify that its orbit is not affected by two-body relaxation, \cite{Amaro-Seoane2007} define a critical semi-major axis below which GW emission dominates orbital evolution as:

\begin{equation}
\label{a_emri}
a_\mathrm{EMRI} =  5.3 \times 10^{-2} \mathrm{pc}\times C^{2/3}_\mathrm{EMRI} \left(\frac{t_\mathrm{rc}}{Gyr}\right)^{2/3}\left(\frac{m}{10M_\odot}\right)^{2/3}\left(\frac{M_\mathrm{SMBH}}{10^6M_\odot}\right)^{-1/3}
\end{equation}
where $C_\mathrm{EMRI} \simeq 1$ and $t_\mathrm{rc}$ is the local relaxation time. The latter is given by \citep{Spitzer1987, BinneyTremaine2008}:

\begin{equation}
\label{t_rc}
t_\mathrm{rc} =  \frac{18\mathrm{Gyr}}{\ln\Lambda}\frac{1M_\odot}{m_\mathrm{bh}}\frac{10^3M_\odot \mathrm{pc}^{-3}}{\rho(r)}\left(\frac{\sigma(r)}{10\mathrm{km/s}}\right)^3, 
\end{equation}
where $\ln\Lambda$ is Coulomb logarithm, $m_\mathrm{bh}$ is the mass of a stellar BH, $\rho$ and $\sigma$ are the stellar density and 1D velocity dispersion respectively. Assuming $\ln\Lambda \simeq 10$, $\sigma = \sqrt{GM_\mathrm{SMBH}/3r}$, measuring the BH mass to be $m_\mathrm{bh}$$= 10$ $M_\odot$ and the density of stellar BHs $\rho=4\times10^4$ $M_\odot \mathrm{pc}^{-3}$ (see Fig.~\ref{fig:SCDens}), we get $t_\mathrm{rc}\simeq 6.2$ Gyr at $r=0.5$ pc. Thus, for a typical BH in our simulation with $m=10$ $M_\odot$, the critical semi-major axis equals to $a_\mathrm{EMRI} = 0.11$ pc. The late evolution of an EMRI is determined by GW emission, leading eventually to the SMBH - compact remnant coalescence over the merging time due to the gravitational radiation given by \citep{Peters1964} 
\begin{equation}
\label{t_gw}
t_\mathrm{GW} \approx \frac{768}{425} \frac{5}{256}\frac{c^5}{G^3}\frac{a^4}{m_1 m_2(m_1+m_2)}\left(1-e^2\right)^{7/2}
\end{equation}
(here we use an approximation where $e\sim 1$). If we take the eccentricity to be $e=0.9999$, then for a BH with typical mass $m=10$ $M_\odot$ and the critical semi-major axis $a = 0.11$ pc orbiting the SMBH of $4.3\times 10^6$$\msun$ the merging time equals 94 Myr which is much shorter than the relaxation time. The chosen eccentricity corresponds to the pericentre distance $r_\mathrm{peri} \simeq 27 R_\mathrm{S}$. Thus, we define the criterion for an EMRI as:

\begin{eqnarray}
\label{emri_cond}
a  <   a_\mathrm{EMRI};\nonumber\\
4 R_\mathrm{S} < r_\mathrm{peri} < 27 R_\mathrm{S}
\end{eqnarray}

Now we can use the classical loss-cone theory to calculate the number of EMRIs. First, we obtain amount of `loss-cone' orbits with $r_\mathrm{peri} < 4 R_\mathrm{S}$ then with $r_\mathrm{peri} < 27 R_\mathrm{S}$ and take the difference between former and latter, finally we exclude the objects with $a > a_\mathrm{EMRI}$. The same procedure can be applied to calculate number of EMRIs originating from a WD - SMBH binary coalescence. In this case $a_\mathrm{EMRI} = 0.02$ pc and  $r_\mathrm{peri} \simeq 5R_\mathrm{S}$ (we used mean mass of accreted WD from our simulation $m_\mathrm{WD}$$ = 0.7$$\msun$). The last column of Table~\ref{StProp} provides the EMRI rates of compact objects in our simulation. Our results suggest that $\simeq 2$ WDs and $\simeq 2$ BHs are expected to undergo an EMRI over each Gyr. The derived EMRI rates are lower than the previous estimations \citep[e.g.][]{Hopman2009a,Arca-Sedda&Gualandris2018}, likely due to the low fraction of retained NSs and BHs in our model. We can conclude that WDs are the main sources of direct plunges in the GC with a minimum rate of more than 250 events per Gyr, we found few BHs and no NS.  

We note that here we do not follow the accreted object after it is gone inside $4 R_\mathrm{S}$ or classified as an EMRI.

%###########################################################################
\section{Binaries}
\label{sec:bin}
%###########################################################################

\begin{figure}
\begin{centering}
\includegraphics[width=\columnwidth]{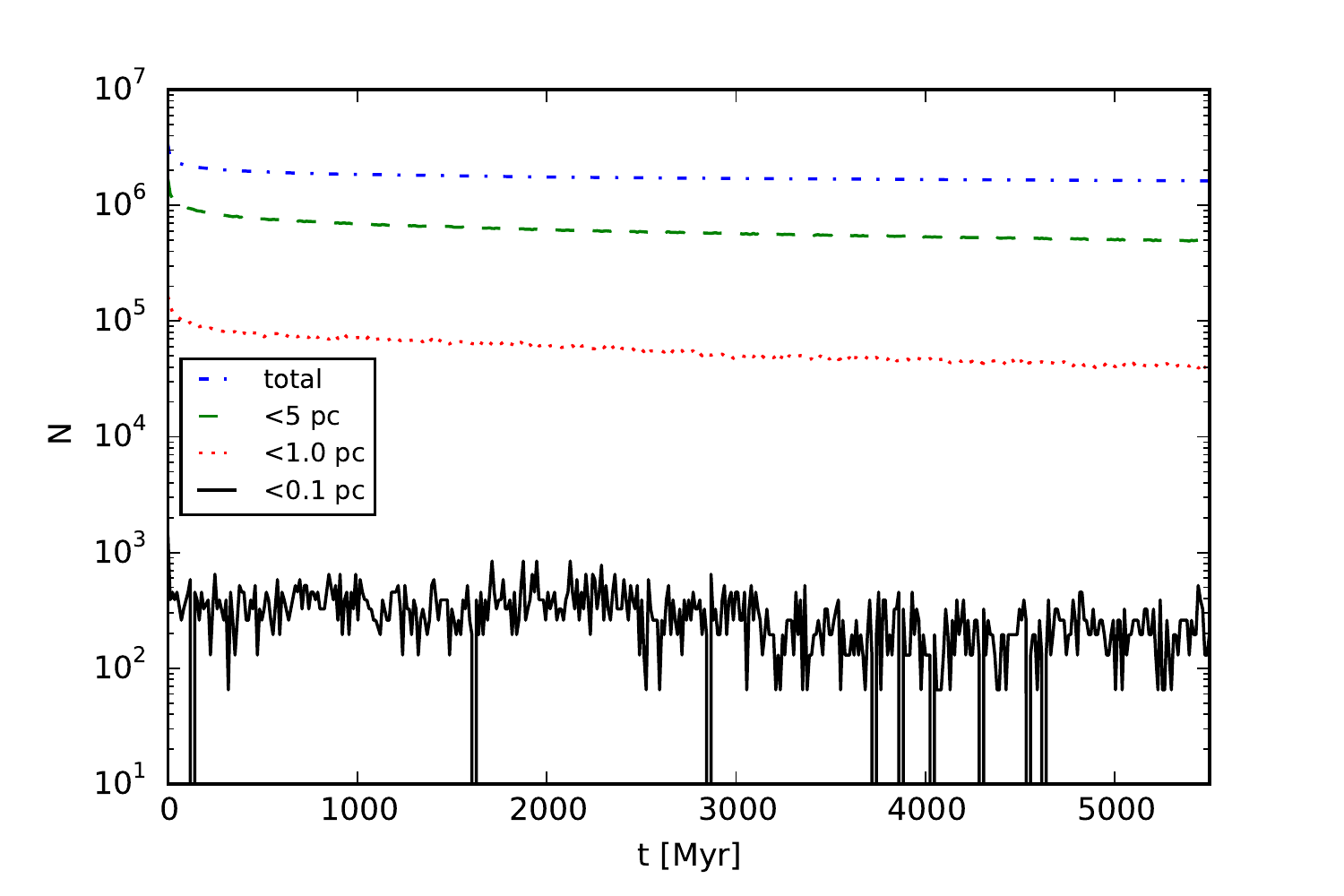} \\ 
\par\end{centering}
\caption{Number of binaries as function of time. The blue line shows the total number of binaries. Cyan, red and green lines show the number of binaries inside 0.1, 1 and 5 pc correspondingly.}
\label{fig:nbin}
\end{figure}

\begin{figure}
	\begin{centering}
		\includegraphics[width=\columnwidth]{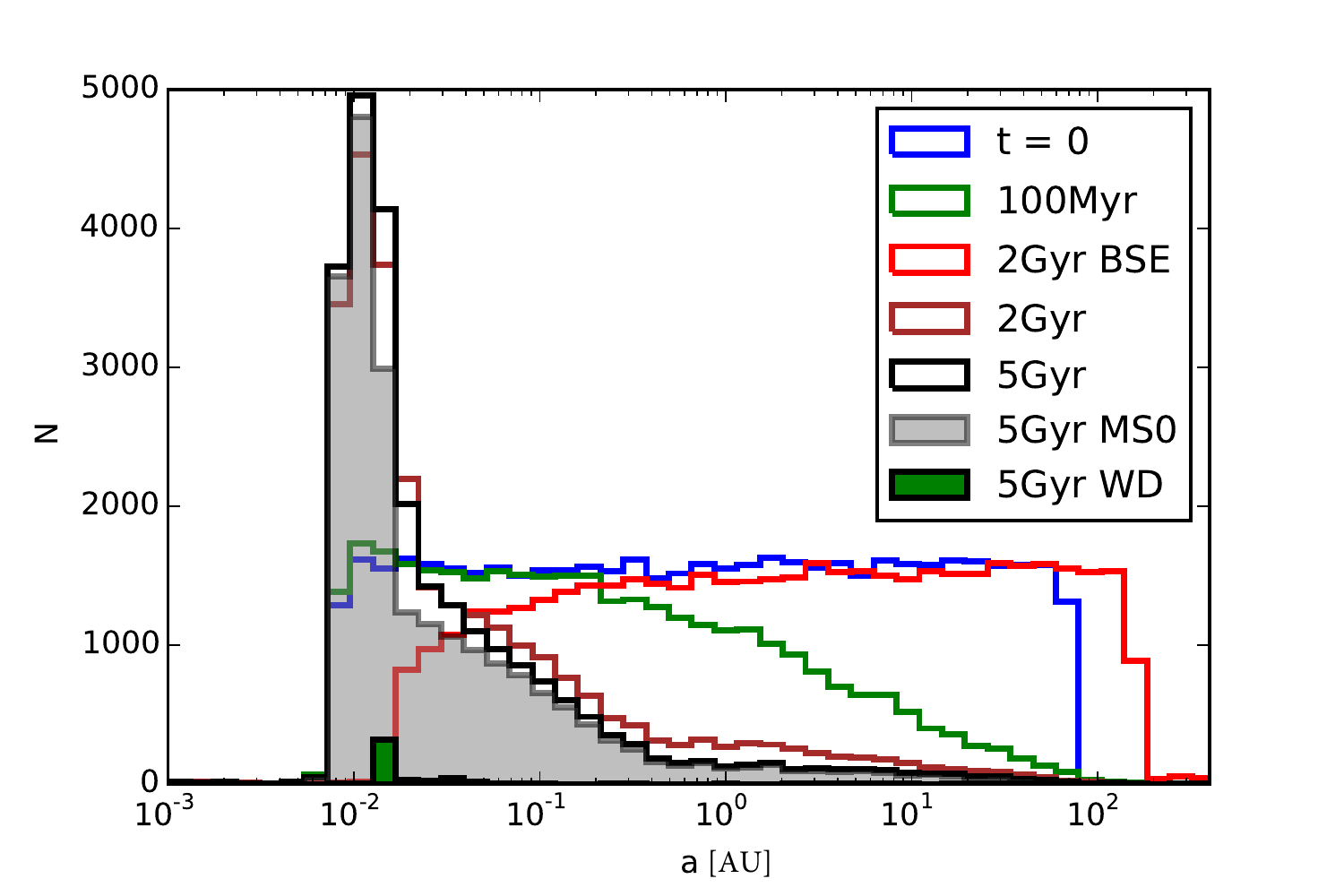} \\ 
		\par\end{centering}
	\caption{Semi-major axis of binaries (not rescaled). The blue line shows the initial distribution, green and black lines represent the semi-major axis at t = 100 Myr and t = 5 Gyr respectively. The red line displays the semi-major axis without dynamics (standalone binary stellar evolution (BSE) simulation). Grey and green step-filled histogram indicate the distribution of double low-mass MS binaries and stellar systems with WDs at 5 Gyr respectively.}
	\label{fig:semi}
	
\end{figure}

Although the GC environment is very extreme, some binaries have been detected there \citep{Muno2005b,Pfuhl2014}. In this section we analyse the number of GC binaries obtained from our simulation. The upper curve in Fig.~\ref{fig:nbin} shows the total number of binaries as function of time. We start the simulation with 5\% of binaries and roughly half of them survive till $t = 5$ Gyr of evolution. The two lower curves in the same figure show the number of binaries inside 1 and 0.1 pc, respectively, meaning that after 5 Gyr we expect 100-1000 of them inside 0.1 pc and $\simeq 5.0\times 10^4$ in the inner parsec. These binaries are characterized by an average total mass of $1.0$ and $0.69 \msun$ and a binary fraction of $\sim 2$\% and $\sim 2.5$\% respectively. The initial distribution of the binary semi-major axis (SMA) was assumed log-uniform between a = 0.005 and 50 AU.
The SMA defines the binary binding energy ($E_b \propto 1/a$), and allows to determine whether a binary is `hard' or `soft' \citep{Heggie1975}. A sizeable number of `soft' binaries are quickly destroyed because of the repeated interactions with the surrounding dense environment, leading to a strong decrease of the number of binary systems having initial SMA values larger than 1 AU (compare blue and green lines in Fig.~\ref{fig:semi}).

\begin{figure}
	\begin{centering}$\begin{array}{c}
		\multicolumn{1}{l}{\mbox{(a) $t = 100$ Myr  }}\\
		\includegraphics[width=0.8\columnwidth]{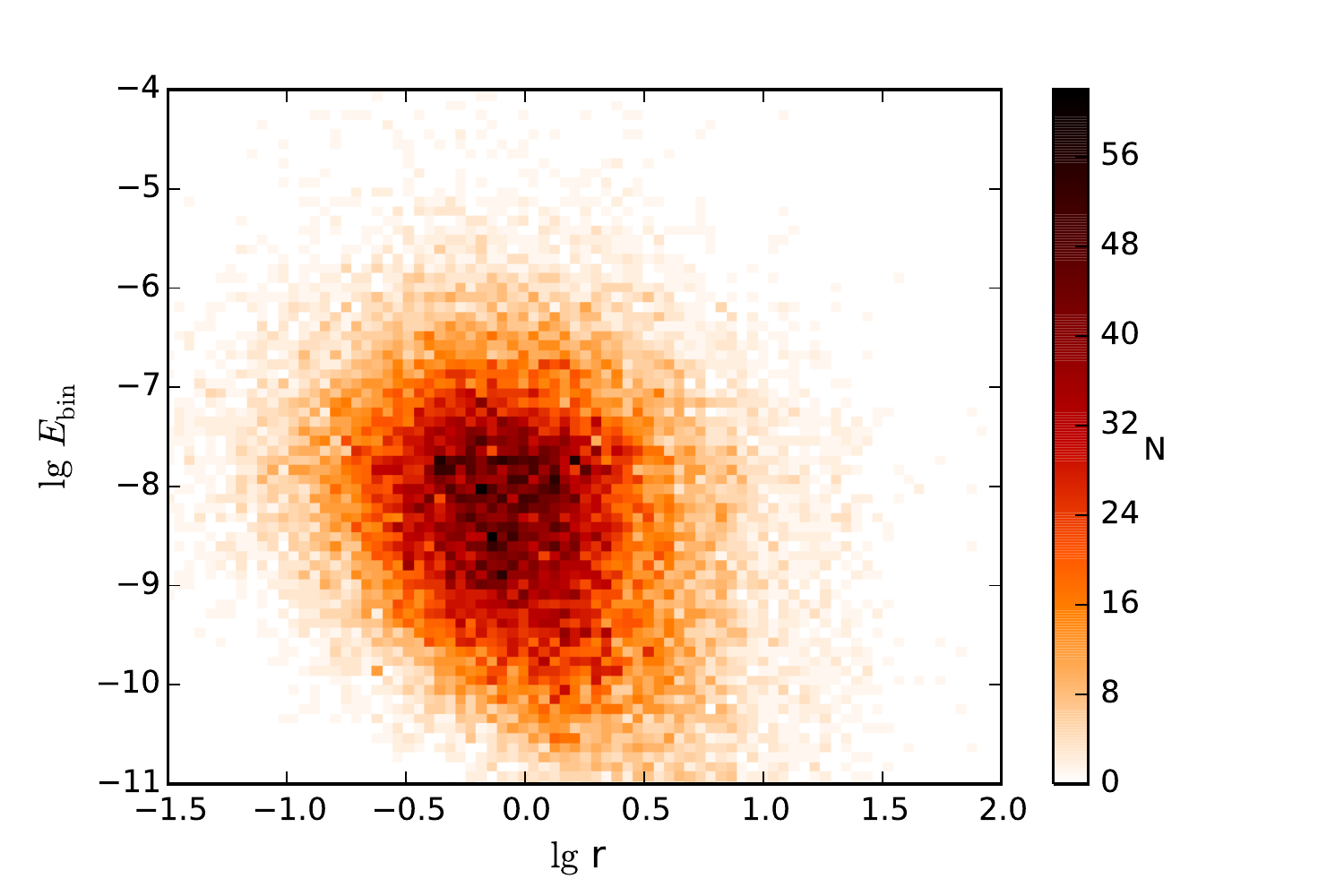}\\
		\multicolumn{1}{c}{\mbox{}}\\
		\multicolumn{1}{l}{\mbox{(b)  $t = 1$ Gyr }}\\
		\includegraphics[width=0.8\columnwidth]{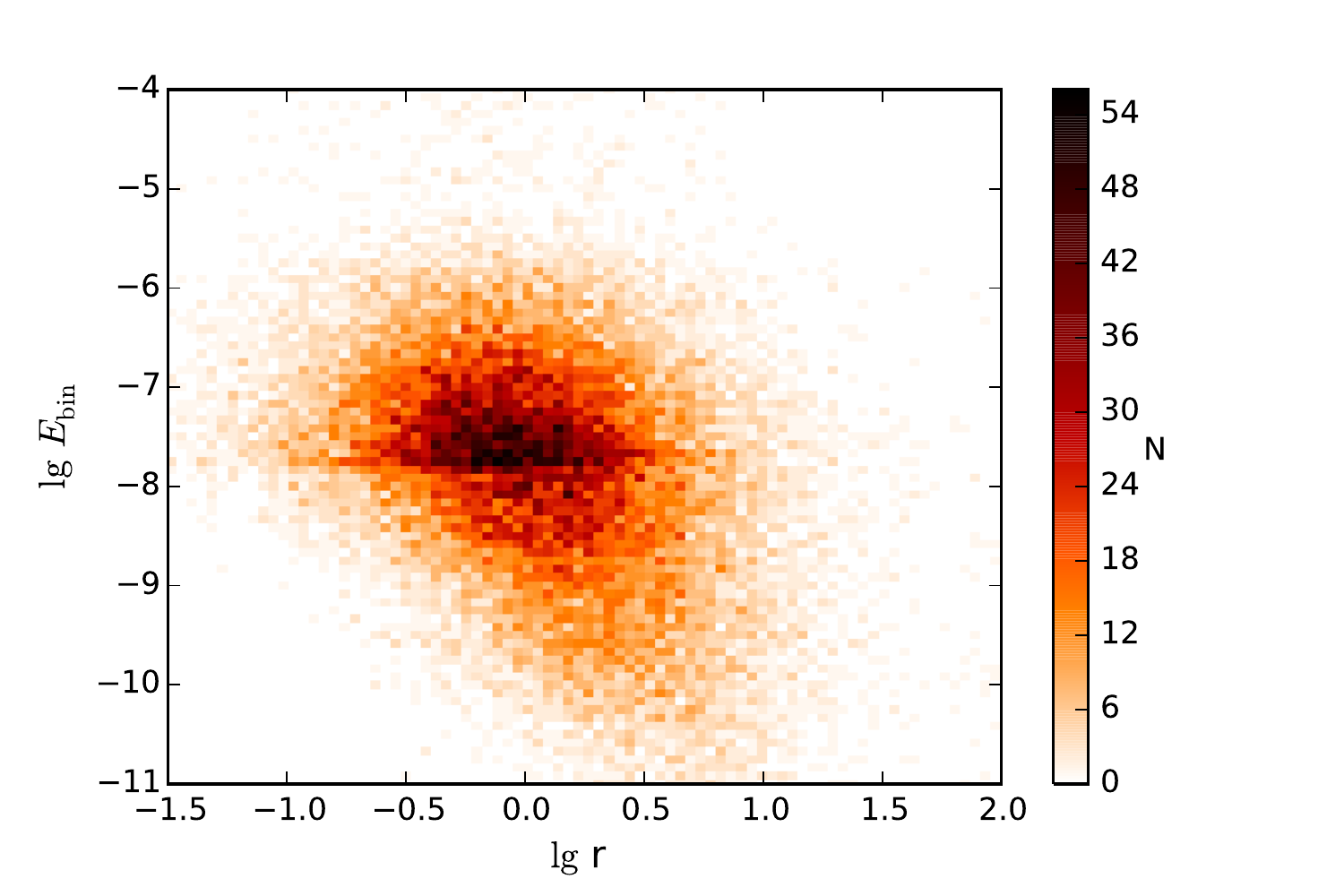}\\
		\multicolumn{1}{l}{\mbox{(c)  $t = 5$ Gyr }}\\
		\includegraphics[width=0.8\columnwidth]{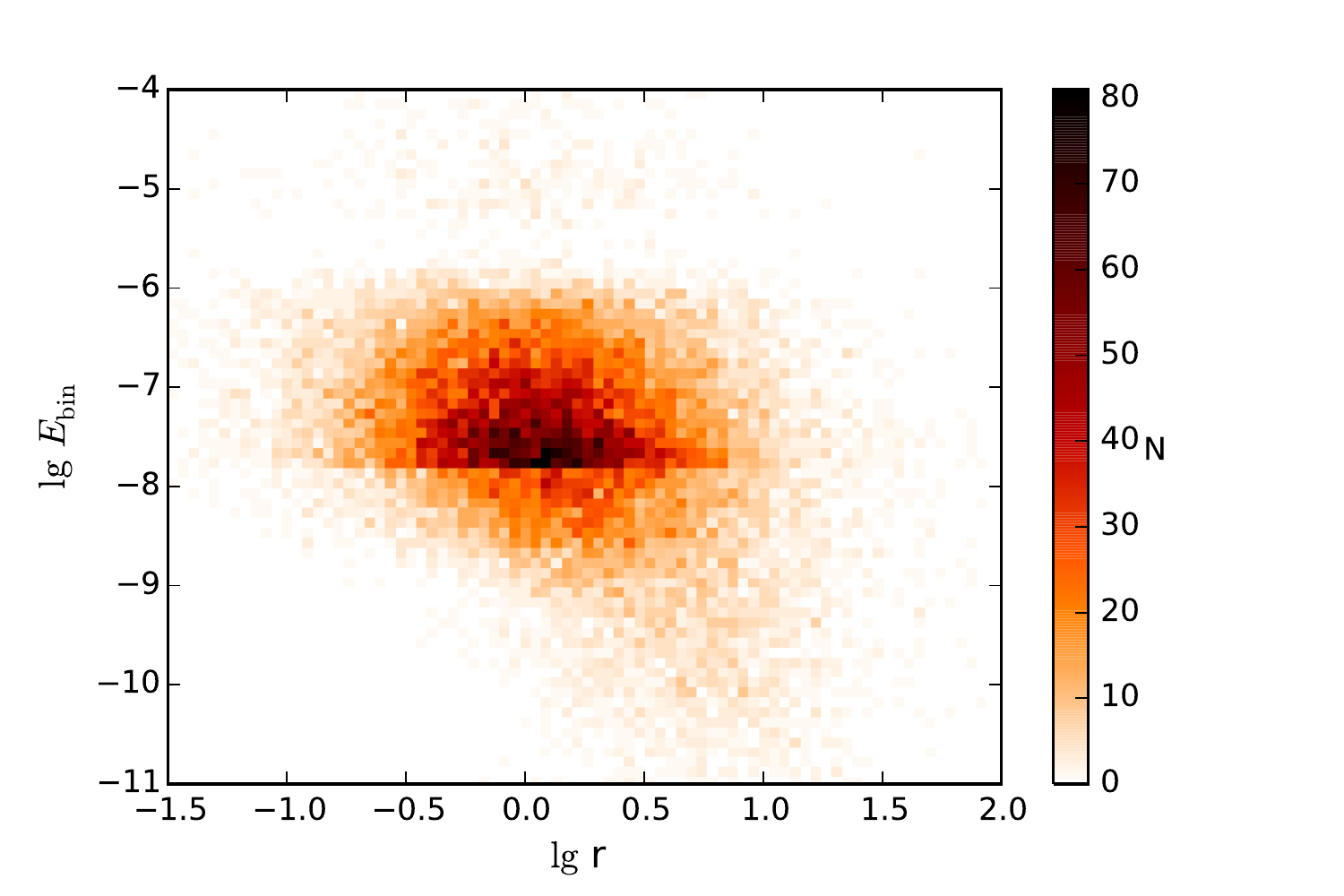}\\
		\end{array}$
		\par\end{centering}
	\caption{2D histogram of binding energies of binaries versus their distances to the SMBH (all values are given in $N$-body units), colour-coding shows the number of binaries in each bin. Panels a), b) and c) correspond to 100 Myr, 1 Gyr and 5 Gyr.}
	\label{fig:ebin_hist}
\end{figure}

On the other hand, the number of systems with smaller SMA increase in time, thus implying a growing number of `hard' binaries. Fig.~\ref{fig:semi} compares the SMA distribution at a time t = 2 Gyr in our simulation (the peaked brown line) with the 
SMA distribution obtained evolving all the binaries in isolation. This comparison shows the effect of the dense environment on the binary stellar evolution. Thus, the systems with small separations are getting higher in number and their orbits shrink. As opposite to this, the standalone binary evolution code results show that the number of binary systems with smaller separations will decrease (some of them will merge and some of them will get wider orbits after the supernovae explosions). Step-filled histograms on Fig.~\ref{fig:semi} show that the low-separation binaries are dominated by low-mass MS stars and some WDs. Fig.~\ref{fig:ebin_hist} shows the distribution of the binding energies of the binaries and their distances to the SMBH at 100 Myr, 1 Gyr and 5 Gyr. As we can see, the number of binaries with binding energies below $10^{-8}$ $N$-body units decreases with time, especially in the central part. 

\begin{figure}
	\begin{centering}
		\includegraphics[width=\columnwidth]{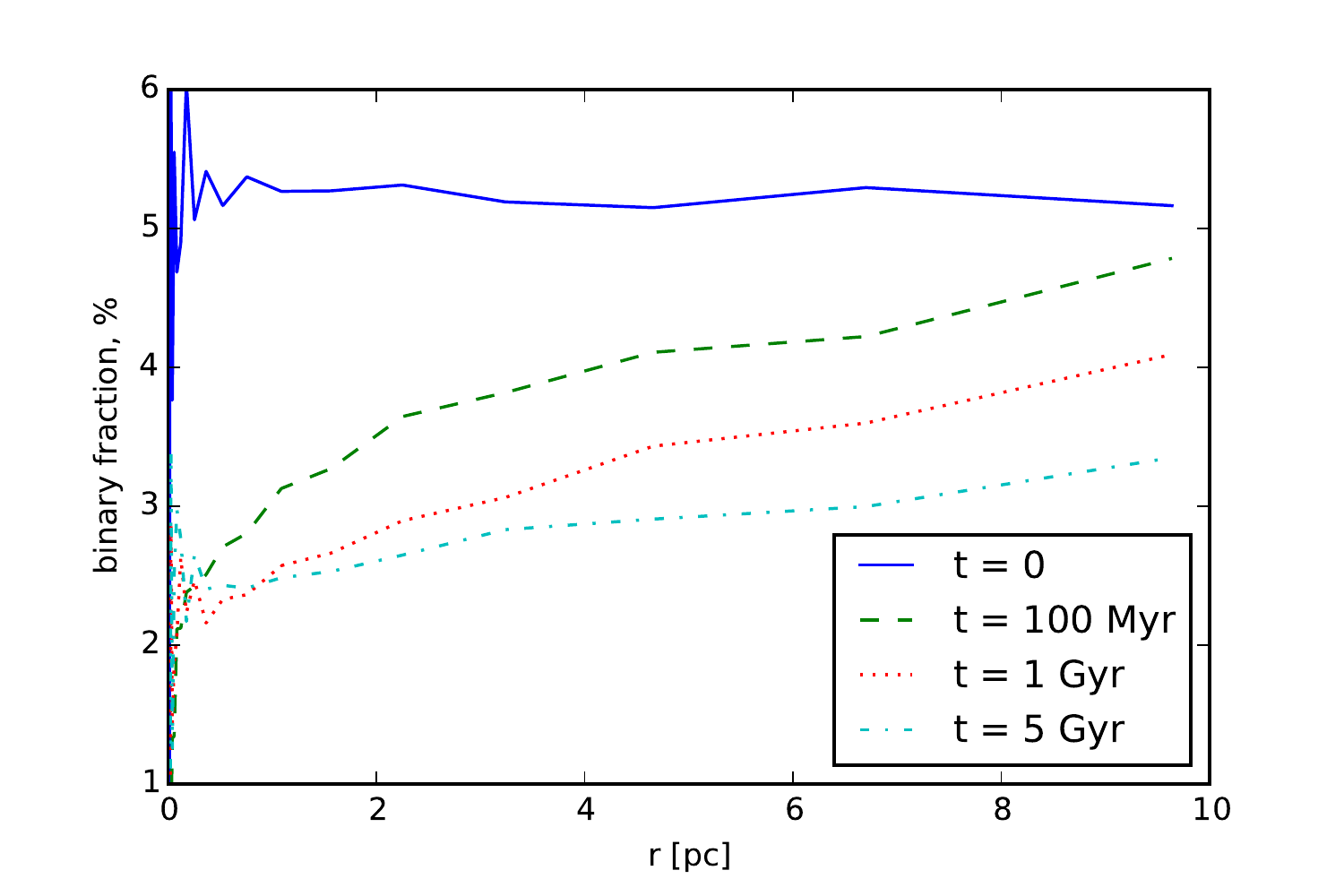} \\ 
		\par\end{centering}
	\caption{Binary fraction as a function of the distance from the SMBH. Blue line shows the initial dependence, green, red and cyan lines represent the binary fraction as a function of radius at 100 Myr, 1 Gyr and 5 Gyr respectively. }
	\label{fig:binfrac}
	
\end{figure}

The total mass of a binary system is typically twice larger than that of a single star, thus implying that most of the binaries will be subjected to mass segregation. While mass segregation brings the binaries to the centre, the soft ones are being destroyed and hard ones survive, but even a very hard binary can be tidally disrupted by the SMBH. Fig.~\ref{fig:binfrac} illustrates how the binary fraction changes with the distance from the SMBH for initial moment (blue curve), 100 Myr (green), 1 Gyr (red) and 5 Gyr (cyan). Initially, the binaries were distributed uniformly but already after 100 Myrs the central binary fraction ($r < 1$ $\mathrm{pc}$) dropped from 5\% to $\sim 2.5$ \%. Comparison of the red and cyan lines yields that the total number of binaries drops but in general the shape of the curve is established.   

\begin{figure}
	\begin{centering}
		\includegraphics[width=\columnwidth]{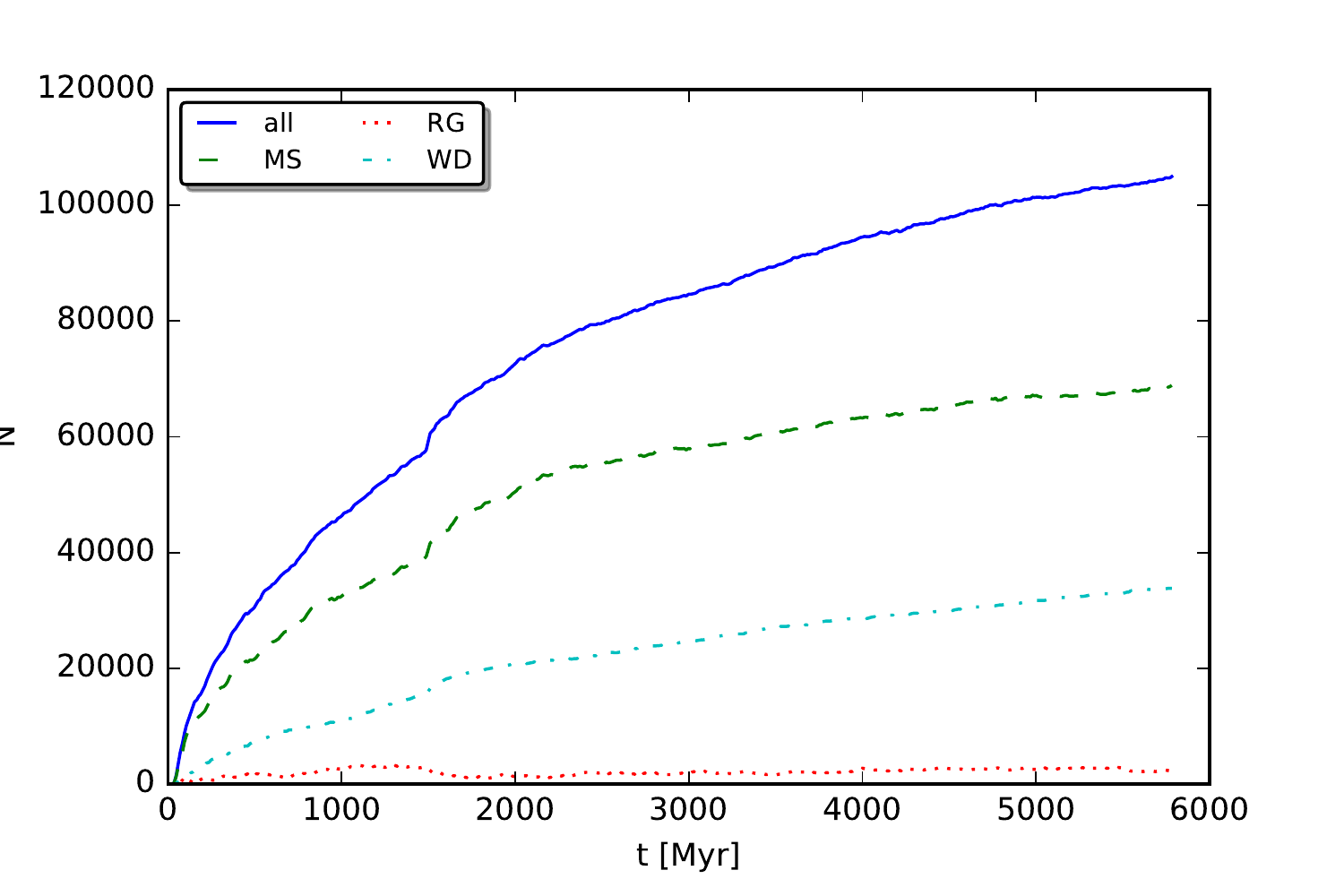} \\ 
		\par\end{centering}
	\caption{Number of WD binaries as a function of time. The blue line shows the total number of WD binaries, green and red lines are WD - MS star and WD - WD binaries correspondingly. }
	\label{fig:wdb}
	
\end{figure}

\begin{figure}
	\begin{centering}
		\includegraphics[width=\columnwidth]{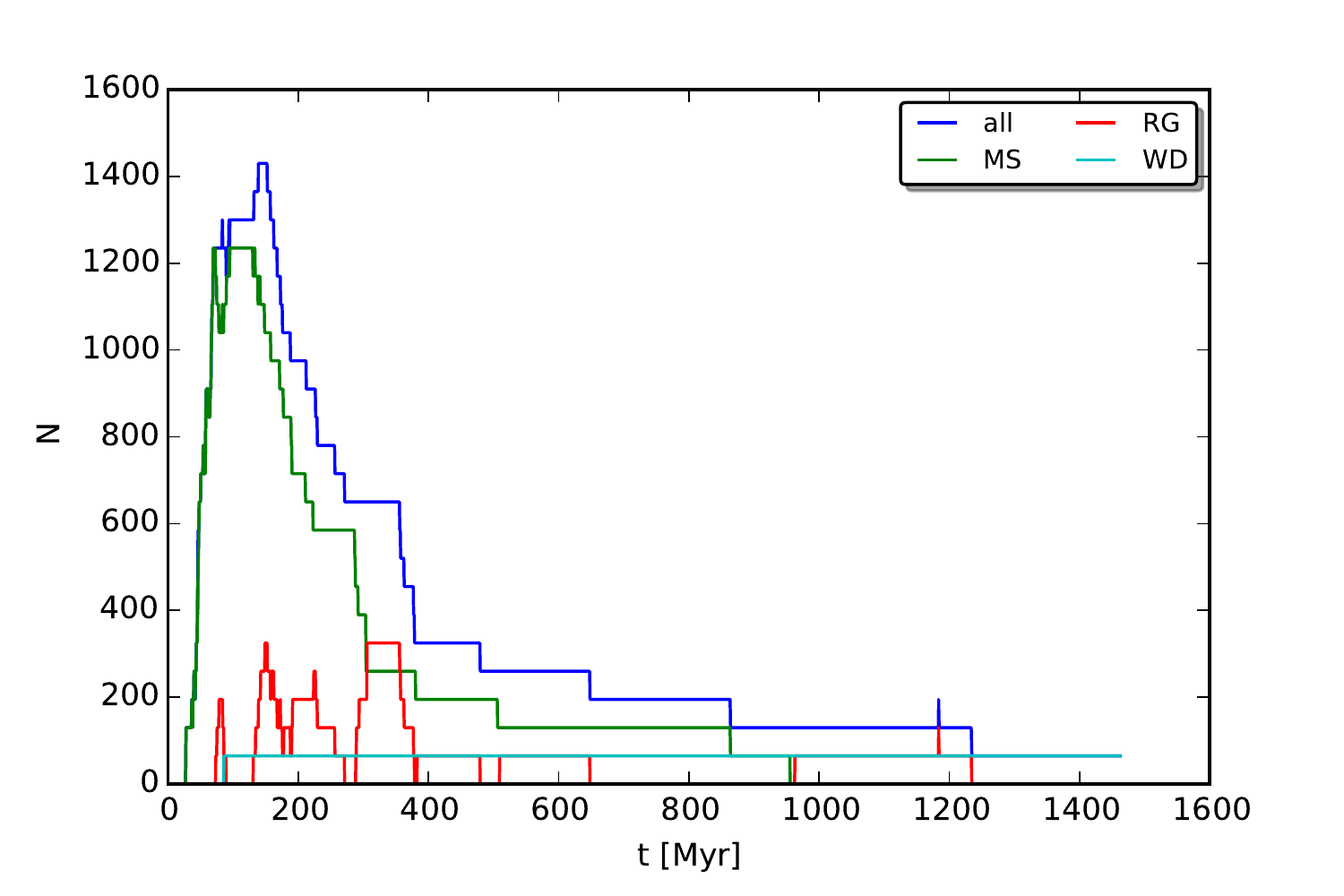} \\ 
		\par\end{centering}
	\caption{Number of NS binaries as a function of time. Blue line: all NS binaries, green: NS - MS star, red: NS - RG and cyan: NS - WD.}
	\label{fig:nsb}
	
\end{figure}

Our simulation suggests that the NSC contain a substantial number of WD binaries (Fig.~\ref{fig:wdb}). These binary systems are of particular interest since they can give rise to supernovae Ia events or, in some cases, they can even form a millisecond pulsar (MSP) through  
matter accretion from a companion star onto a highly spinning massive WD \citep{FreireTauris2014}.
Double degenerate WD - WD binaries can be the progenitor of supernovae Ia explosions, provided that their total mass exceeds the Chandrasekhar limit (note that there are also sub and super Chandrasekhar models, see review by \citealt{Maoz2014}).
On the other hand, binary systems containing a NS
are almost absent in the system (Fig.~\ref{fig:nsb}). These types of binaries are possible progenitors of MSPs, which are thought to be recycled NSs spun up by matter accretion from a stellar companion, according to the standard scenario. After the natal kick the NS binaries become very wide (if they survive the supernova explosion) and eventually are ionized. We find that 1000 and $\approx 3000$ of double and single degenerate pairs are expected to populate the central parsec of the MW galactic nucleus.

%###########################################################################
\section{Summary and discussion} 
\label{sec:CON}
%###########################################################################

We performed a high resolution direct $N$-body simulation of the GC starting with $\sim 10^6$ particles with 5\% of initial binaries taking into account single and binary stellar evolution. This is the largest simulation of this kind so far. We showed that the stellar component forms a cusp with the highest power-law index for the stellar mass BHs $\gamma \simeq -1.72$. Then we demonstrated how mass segregation occurs by analysing average masses between Lagrangian shells. When the stars happen to come very close to the SMBH they are disrupted with a total rate of $\sim 4\times10^3$ stars per Gyr. The number of accretion events for compact objects is $\sim 270$ per Gyr with few of them being possible EMRIs. About half of the initial binaries survived until 5 Gyr of the evolution. Most of the binaries are destroyed due to dynamical interactions with single stars. The increasing number of WD binaries could imply a high supernovae Ia rate. Once the rotation and non-sphericity of the NSC is taken into account, the amount of TDEs/EMRIs could increase, and the evolution of the system may differ. These effects are to be studied in follow-up papers.

The absence of NSs in the NSC after ~200 Myr of evolution is due to high velocity kicks (the velocity distribution of such kicks is still highly debated, see e.g. \citealt{Beniamini2016}) at the moment of formation of a NS. This lets them escape from the system. In case of a binary, if the latter survives the supernova explosion, the binary gets a very wide orbit and eventually is destroyed by interaction with single stars in the dense stellar environment. Thus, postulating that all NSs have velocity kicks with 1D dispersion of $265\kms$, makes the formation of a close binary with a NS unlikely. Therefore, the standard scenario for formation of MSPs fails due to a lack of NSs. In this sense our results are consistent with the more simplistic approach by \cite{Bortolas2017} where the authors claim that most of the NSs progenitor binaries do not survive the supernova explosion. But in reality MSPs are observed even in globular clusters \citep[e.g.][]{Manchester1991} where the escape velocity is much smaller than for a NSC. It means that MSP can form in an alternative scenario, for example from the accretion induced collapse of a WD \cite[e.g.][]{Hurley2010, Taani2012, Tauris2013, FreireTauris2014}. If a MSP is detected in the close vicinity of a SMBH it can be used to test general relativity in the strong regime \cite[e.g.][]{Psaltis2016}. Moreover, the spatial distribution of MSPs in the NSC can give hints on the formation scenario of the NSC \citep{Arca-Sedda2017, Abbate2018}. We note that the estimation of number of MSPs in the GC is still to be analysed in more detail. We aim to start several new runs taking into account the MW bulge as an external potential and investigate how many NSs would be bound to the NSC. We expect that the bulge will prevent NSs from escaping and lead to an increase in number of progenitors of MSPs.  

A 3-body interactions involving a binary star and the SMBH can result in the binary break up, with one component being captured by the SMBH and the other ejected away with velocities up to 1000 $\kms$ \citep{Hills1988}. This mechanism is one of the possible scenarios that can explain the observed population of hypervelocity stars \citep{Brown2015}. Indeed, if a binary with mass 1 M$_\odot$ and semi-major axis of 0.1 AU approaches to the MW SMBH as close as its disruption radius $ r_\mathrm{bt}\simeq 10^{-6}$ pc, then it may lead to the formation of a hypervelocity star with $v\simeq1370 \kms$ (Eq. 2 in \citealt{Brown2015}). As we have seen from Fig.~\ref{fig:semi}, the NSC is completely dominated by binaries with small separations at 5 Gyr, meaning that it is likely to expect hypervelocity stars with velocities above $1000 \kms$. Our results suggest the majority of the ejected objects are low-mass MS stars or, more rarely, WDs. Since the accretion radius in our simulation exceeds $r_\mathrm{bt}$ for most of the remained binaries, we leave the detailed analysis of the 3-body interactions that potentially involve them and the SMBH to a forthcoming work.

In this simulation we constructed the initial conditions assuming the in-situ formation of the NSC, but its formation is likely due to star cluster inspiral, at least in part, as firstly suggested by \cite{Tremaine1976} and
\cite{Capuzzo-Dolcetta1993}, although a fraction is likely due to in-situ star formation \citep{King2003, King2005}.
\citet{Antonini2012} provided the first self-consistent simulation tailored to reproduce the MW observational properties. 
Later on, \citet{Arca-Sedda2015} showed that the formation of a NSC around an SMBH weighing a few $10^6$ $M_{\odot}$ is
extremely rapid, lasting 0.1-1 Gyr, thus implying that the contributing clusters still are ``dynamically young" when arrive to the GC.
Moreover, \citet{Arca-Sedda2015} presented the first simulations to model self consistently a whole galactic nucleus and 11 star clusters using,
for the whole system, more than $10^6$ particles. More recently, \citet{Tsatsi2017} pointed out that the MW NSC rotation can be reproduced by the ``star cluster inspiral" scenario. Taking into account these facts, our follow-up simulations may be started with the initial stellar distribution according to the ``star cluster inspiral" scenario with some initial rotation.

We note that the rate of TDEs may be enhanced in the presence of an accretion disc \citep{JustEtAl2012,KenEtAl2016}. The same is true for the gravitational waves: the drag force of the accretion disc may bring compact objects close to the SMBH resulting in the enhancement of the EMRI rates detectable by LISA, moreover, the gaseous disc may significantly reduce the SMA of stellar binary BHs boosting their merging time \citep{Bartos2017,Stone2017,McKernan2017}. In case of NS or WD binaries this mechanism may lead to an enhanced rate of supernovae Ia explosions and gamma-ray bursts. Stellar binaries may merge in the close vicinity of the SMBH due to ``eccentric Kozai-Lidov" mechanism  \citep{AntoniniPerets2012, Prodan2015, Stephan2016}. The Kozai-Lidov oscillations can be studied via the direct $N$-body modelling with one-to-one particle resolution or by approximating outer stars as a smooth potential. \citet{Panamarev2018} showed that the interaction of stars with the accretion disc may lead to formation of a nuclear stellar disc in the inner part of the galactic nucleus. Such stellar discs may serve as environment for dynamical formation of compact binaries. 

%###########################################################################
\section{Acknowledgements}
%###########################################################################

The authors thank the anonymous referee for useful comments
and suggestions, and Ortwin Gerhard, Stefan Gillessen and Chengmin
Zhang for very helpful personal discussions about an early draft
of this paper. 

TP acknowledges the support within the program BR05336383 funded by the
Ministry of Defense and Aerospace Industry of the Republic of Kazakhstan.

We thank the Excellence Initiative
at the University of Heidelberg, which supported us by
measures for international research collaborations.

This work was partly supported by Sonderforschungsbereich SFB 881 
``The Milky Way System" (subproject Z2) of the German Research Foundation (DFG).
MAS also acknowledges
financial support from the Alexander von Humboldt Foundation
and the Federal Ministry for Education and Research
in the framework of the research project ``The evolution of
black holes from stellar to galactic scales". LW acknowledges financial support from the Alexander von Humboldt Foundation.

This work has been supported partly by the
National Science Foundation of China (NSFC) under grant No. 11673032,
by Strategic Priority Research
Program (Pilot B) ``Multi-wavelength gravitational
wave universe" of the Chinese Academy of Sciences (No.
XDB23040100), and by
the China-Kazakhstan international cooperation project of Chinese 
Academy of Sciences (CAS) "Evolution of star clusters in galactic
nuclei with supermassive black holes and central accretion
disks" under the project number 114A11KYSB20170015.
PB has been supported by President's International Fellowship 
for Visiting Scientists program of CAS. 

We acknowledge the support of the Volkswagen Foundation (VW)
under the Trilateral Partnerships grant No. 90411, 
and through using the kepler computer at ARI Heidelberg, 
funded by the project GRACE 2: ``Scientific simulations
using programmable hardware" (VW grants I84678/84680).
The authors gratefully acknowledge the Gauss Centre for Supercomputing (GSC) e.V. 
(www.gauss-centre.eu) for funding this project by providing computing time 
through the John von Neumann Institute for Computing (NIC) on the 
GCS Supercomputers JURECA and JUWELS at J\"ulich Supercomputing Centre (JSC).

Special support by the National Academy of Sciences of the Ukraine (NASU) is
provided under the Main Astronomical
Observatory GRID/GPU computing cluster project.
This work benefited from support by the International
Space Science Institute, Bern, Switzerland, through its 
International Team programme ref. no. 393 ``The Evolution of
Rich Stellar Populations and BH Binaries" (2017-18).

\bibliography{GCrefs}

\end{document}